# Reforming the State-Based Forward Guidance through Wage Growth Rate Threshold: Evidence from FRB/US Simulations[1]

Author: Sudiksha Joshi, Summer Data Analyst at Employ America,

joshi@employamerica.org

**Abstract**

I have analyzed the evolution of forward guidance from calendar to state-based, the practicality of the Evan's Rule and possible ways to reform it. The Evan's Rule prescribes that the liftoff of the federal funds rate from the zero lower bound (ZLB) until the unemployment rate and/or the PCE inflation rate thresholds are breached. In the process, I examined the biases, measurement errors, and other limitations extant in the labor utilization and inflation variables, and reviewed the literature on methods to improve forward guidance, particularly when the federal funds rate is at ZLB. Using time-series analysis−after testing for stationarity and cointegration, I calibrated the thresholds of ECI wage growth and the employment-to-population ratio and investigated the relationship between other labor utilization variables. Then I imposed various supply, demand and monetary shocks and constructed impulse response functions to contrast the paths of eight macroeconomic variables including the federal funds rate, 10-year treasury yield, GDP growth rate under three scenarios−consensus baseline, current Evan's Rule and the new ECI wage rate threshold. The results suggest that under the wage growth rate scenario, the federal funds rate lift off earlier than under the current Evan's Rule, indicating a tightening of the monetary policy.

[1] I gratefully acknowledge the advice from Skanda Amarnath, the Director of Research and Analysis at Employ America, throughout this project. All errors (if any) are my own. For discussion, feedback or comments, please email me at joshi@employamerica.org.

# Introduction

In this research paper, I have empirically examined the feasibility of state-based forward guidance in the US, specifically the role of Evan's Rule in shaping the economy, and constructed FRB/US simulations to explore alternative ways the FOMC can employ to enhance state-based forward guidance. The FOMC employs the unconventional monetary policy tool of forward guidance so that the public can understand the expected path of interest rates. Particularly at the Zero Lower Bound (ZLB), it amplifies monetary policy accommodation. By assuring the financial markets and the public, it reduces the uncertainty about future policy. Forward guidance and large scale asset purchases (first introduced on November 25, 2008) impose downward pressure on the long term yields. Whilst large scale asset purchases (LSAPs) reduce the term premium part of the long term yields, forward guidance lowers the investors' future expectations of short-run yields.

On December 12, 2012, when the FOMC extended to purchase longer-term assets under LSAP3, it switched its forward guidance formulation from calendar dates to threshold-based or state-contingent forward guidance. It linked the forward guidance on policy rates directly to its economic objectives. Unlike in its previous post-meeting statements wherein it declared to keep the federal funds rate low at least until a specific year, now the FOMC introduced thresholds of unemployment and the projected PCE inflation rate (one to two years from the present) at 6.5 and 2.5 percent, respectively, that is the foundation of the Evan's Rule. However, breaching these thresholds would not automatically signal an increase from the current level of the rates held at 0 to 25 bps. Further, it clarified that the rise of the federal funds rate from the ZLB is not entirely contingent upon crossing the thresholds as long term inflation expectations have to be well-anchored. Moreover, the guidance was based on thresholds, not triggers. Upon crossing the threshold, the members of the FOMC would consider a range of other variables that affect the health of the labor market to check if policy rate hikes are warranted.

The structure of the research paper is as follows. First, I've reviewed the literature on forward guidance, then discussed the deficiencies of adopting the Evan's Rule with the unemployment and inflation rate thresholds of 6.5 and 2.5 percent, respectively. Focusing on the time period 2012-2017 when the FOMC had established the Evan's Rule as a framework to conduct forward guidance, I calibrated the threshold values of employment to population ratio and ECI wage rate, which are comparable to the 6.5 percent unemployment rate, and 2.5 percent PCE inflation rate. Using the threshold value of wage growth, I simulated the impact of supply, demand and monetary shocks on macroeconomic variables under two scenarios − first, in which the FOMC adheres to the currently established Evan's Rule, and, second, in which it keeps the unemployment rate threshold at 6.5 percent, but replaces the PCE inflation rate with the ECI wage growth of 3.5 percent. Additionally, I have constructed the aforementioned simulations for the years 2020-Q2 to 2023-Q4. Besides the thresholds, I've altered the parameters in the optimal control policy and the Taylor rule monetary policy reaction function to gauge if changing weights on output and inflation gap impact the timing of the liftoff of the federal funds rate from the zero lower bound (ZLB).

# Literature Review

Raskin (2013) evaluated the effects of calendar-based forward guidance and discerned the degree to which it changed the public's opinions on the FOMC's monetary policy reaction function. He used estimates that measure how sensitive the money market futures rate are to surprises in the macroeconomic data. Albeit, the ZLB on the nominal interest rates complicates the process in addressing the above question, he overcame the problem by constructing risk-neutral probability density functions of the public's expectations on the short-term interest rates, which are derived from interest rates options. The empirical model comprises dummy variables to reflect the impact of the extension of calendar-based forward guidance, keeping other factors constant. These factors are volatility in the monetary policy and business cycle which vary over time when shocks hit the economy. The point estimates from the regression models signal that the phrase "mid-2013" curtailed how sensitive the expectations were by 75 to 100 percent.

In August 2011, when the FOMC revised its guidance to keep the federal funds rate low "at least through mid-2013," investors lowered their expectations on the future federal funds rate target for the next couple of years, declining the money market futures rates. However, the announcement diminished the uncertainty surrounding its path as the implied volatility on options considerably declined. He ascribed this to the increasing probability of ZLB being a binding constraint in the future. He found that the calendar-based forward guidance in August 2011 led to a statistically significant change in the market's perceptions of the reaction function. Consequently, the risk-neutral percentiles became less sensitive to economic shocks. Yet, the FOMC's statement to extend the lowering of the policy rates from "mid-2013" to "late-2014" in January 2012 had no statistically significant effect as the market participants had previously anticipated the FOMC to extend the guidance. As the short-term interest rate expectations became less sensitive, the long-term rates became anchored at lower levels. Furthermore, people's perception of changing reaction function stimulated demand and pushed up consumer confidence.

Distinguishing the effects of LSAPs and forward guidance is fraught with difficulty as FOMC has on several occasions announced both policies simultaneously. Swanson (2017) employed the identifying assumptions in the approach followed by Ganurkaynak, Sack, and Swanson (GSS) method to separately recognize forward guidance from the variations in the federal funds rate. Then he suggested a new identifying restriction to separately identify LSAPs from the other factors. The results of this strategy are robust. Finally, he inputted the options data to estimate how LSAPs and forward guidance modify the uncertainty prevailing in the financial market. Swanson used high−frequency regressions and found that both LSAPs and forward guidance had statistically significant effects on a diverse array of securities such as corporate bonds, treasuries, exchange rates and options that measure interest rate uncertainty. Their effects are similar in magnitude to those produced by the federal funds rate prior to ZLB. While forward guidance influenced short-term treasury yields more than LSAPs, the latter considerably affected the long-term corporate bonds and treasury yields. The results also depicted that LSAP announcements had persistent effects. Even though the impact of forward guidance was less persistent, the difference was statistically insignificant, probably attributed to near-term horizons of forward guidance announcements.

Zhang (2019) calibrated, identified, and separated the two components of the FOMC statements from 2008 to 2015 - LSAPs and forward guidance. The theoretical model highlights that forward guidance announcing easing policies lowers the treasury yields of all maturities. This contrasts with LSAPs that reduce the long and medium-term yields and raise the shortest term treasury yields due to the "feedback of the interest rate rule." The model incorporates various channels for transmitting unconventional monetary policies. Next, she aggregated the impacts of those policies adopted and quantitatively estimated the influence of each FOMC announcement on the real economy and the financial markets. The structural model predicts various responses of the interest rates using the observed high-frequency data. Zhang built a New Keynesian DSGE model and includes a nominal short-term shadow interest rate and a forward guidance shock via FOMC announcements on the future policy rate rule. The model simulations work under ZLB where nominal interest rates remain endogenously when the economy enters a recession. In the wake of a negative shock, the FOMC resorts to forward guidance or purchases long term securities. Then she constructed impulse response functions tracking the movement of interest rates in the event of shocks. In decomposing the Fed's statement into LSAPs and forward guidance, she compared the predicted change in the yield curve from the linear combination of two shocks with the actual change noticed in the yield curve. Then, she estimated a time series each for forward guidance and LSAP announcement and built a structural model to study how persistent these monetary policy shocks were on the aggregate economy. The results showcase that LSAPs were crucial in influencing output whereas forward guidance strongly affected inflation.

Evans et al. (2012) examined the effectiveness of forward guidance as an unconventional monetary policy tool during the Global Financial Crisis (GFC) by simple regression analyses. They also build a Chicago dynamic stochastic general equilibrium (DSGE) model to construct forecasts of macroeconomic variables under different policies and compare the simulation results with the "bright-line economic thresholds." An example of a bright-line threshold is the 7/3 rule wherein the FOMC maintains an exceptionally low level of the policy rate at least as long as the medium-term inflation rate is below 3 percent, and the unemployment rate is higher than 7 percent. Furthermore, they differentiated between the Odyssean and Delphic communications as two ways of conducting forward guidance. In the former, the FOMC publicly commits to hold the future interest rates at a particular range that deviates from the range normally suggested by the underlying monetary policy reaction function. Consequently, it alters the market expectations, measured through the Blue Chip consensus forecasts. In the latter, the FOMC doesn't adhere to a fixed rate and sets the policy rate prescribed by the reaction function when the economic outlook changes. Typically arising during downturns, the Odyssean communications expose deficiencies in a Delphic structure. Resorting to outcome-based policies is paramount in such circumstances and policymakers may have to make hard choices to enforce less understood and unconventional tools to achieve their mandated goals. The FOMC relies on Delphic communication to policies built on the well-established policy framework and consensus forecast inferences in normal times.

On the DSGE models, they investigated the effects of imposing a strong commitment to an accommodative policy in the future and introduced Odyssean forward guidance shocks in the policy rule. They also assume that a monetary policy reaction function derives the path of the federal funds rate when the FOMC makes projections and responds to changing economic conditions over time. Applying the Gurkaynak-Sack-Swanson (GSS) event study technique, they show that forward guidance has significant impacts on asset prices, particularly treasury yields in post-GFC. They administer a sequence of shocks while modeling Odyssean forward guidance for one year from the present. The results indicate that the market expects that the interest rate deviates from the policy rule at least a quarter in advance. Moreover, the markets anticipate 40 percent of the deviations two to four quarters prior.

Charles Evans viewed that threshold-based forward guidance combined with LSAPs enhanced the financial state and real activity as evident from an increased number of purchased durable goods, auto sales, commercial real estate, and housing market data. Offsetting the worldwide and fiscal headwinds faced during and in the aftermath of the GFC, the marginal benefits of threshold-based forward guidance was large given that the US economy was far from reaching its dual mandate objectives. Citing his own original proposal of 7/3 percent - bright-line economic thresholds rule, as conservative due to inflationary pressures, Charles Evans' proposed new thresholds of 6.5/2.5 percent, known as the Evans Rule. He regarded that it would be premature to lower or end the accommodation and its positive effects simply because the unemployment rate hit 6.9 percent - a level greater than the rate linked with maximum employment at that time. We can attribute this reasoning to the macro-model FRB/US simulations wherein keeping the federal funds rate at ZLB until the unemployment rate is at least 6.5 percent causes minimal risks of inflation. Evans initially reckoned the 3 percent inflation threshold as a "symmetric and reasonable treatment" for the 2 percent inflation target and is consistent with the volatility in the inflation rate and the bands of uncertainty surrounding its forecasts. The simulations suggest that the economy is more likely to first cross the unemployment rate threshold of 6.5 percent before the PCE inflation rate spikes to a moderate level of 2.5 percent. Furthermore, he proposed utilizing the forecast of personal consumption expenditures price index, rather than the actual PCE inflation rate to prevent triggering a reaction in the policy rate due to transitory swings in prices of volatile goods such as food, oil, etc. Finally, PCE inflation rate forecasts encompass various economic phenomena ranging from inflationary expectations to cost pressures - these are the variables that affect the inflation rate outlook before they are evident in the actual data on PCE inflation. Therefore, he reckoned the ex-ante measures as a more appropriate safety than a backward-looking measure.

Whilst the prevalent discussion on forward guidance encompasses an inflation "ceiling," Zaman & Knotek (2013) incorporate an inflation "floor." Essentially, an inflation floor commits not to raise the target federal funds rate if the inflation rate lies below a pre-specified value, albeit the unemployment rate crosses its threshold. They use a Bayesian Vector Autoregression (BVAR) to scrutinize when the thresholds of projected inflation and unemployment rates could be breached. It encompasses seven variables that include the federal funds rate, measures of real activity such as real GDP, two measures of inflation - core and total PCE, supply-side variables - unit labor cost, and unemployment rate.

Their model suggests one possible outcome that the unemployment rate would cross the threshold by 2015-Q3 before the actual PCE inflation rate crosses the projected inflation rate threshold. The point forecast of each variable is the mean of different simulated forecast paths of that variable. Besides a point forecast, they predict a joint probability distribution of a spectrum of possible outcomes which display that there is more than 50 percent chance that at least one threshold will be breached in 2015-Q1. The alternative forecasts account for the uncertainty inherent in the modal outlook as they recognize different possible timings when the thresholds might be breached. Post model estimation, they simulate various ex-ante paths of the variables included in the model. Each simulation demonstrates distinct shocks that might occur in the future, alongside the realized values from the estimated distribution of coefficients. Their model depicts that when the inflation rate floor is 1.5 percent, then the point wherein both floors satisfy and at least one threshold crosses - delays by one quarter. This contrasts with an inflation rate floor of 1.75, in which the delay is as long as four quarters. Therefore, the liftoff of the federal funds rate from the ZLB may be substantially delayed depending on the selected rate of the inflation floor.

Next, I have elaborated the reasons that the unemployment rate and inflation rate are inadequate (and possibly inappropriate) variables that formulate the Evan's Rule.

**Biases and Limitations in the Measures of Labor Utilization and Inflation**

The unemployment rate, which is calculated from the data collected in the Consumer Population Survey, is rife with multiple inconsistencies. Firstly, as Hamilton and Ahn (2020) explained, the unemployment rate and the labor force participation rate have rotation bias: in rotation 1, households are visited for the first time, while in rotation 2, they are interviewed for the second time. Likewise, they are interviewed multiple times until the eighth rotation. The BLS uses data from all the eight rotations to calculate the unemployment rate for a specific month. However, the unemployment rate numbers across different rotations vary significantly. This will negatively impact any inference that we conclude from the CPS data. To exemplify, if we collect data from a certain fixed sample of people over periods, then we will notice that the percentage of people employed over time increases i.e. the outflow from unemployment outpaces the inflows.

Secondly, the missing observations documented over time are not random. The frequency at which households in the CPS have been answering all the questions has dwindled. Typically, the BLS calculates the unemployment rate statistics by compiling data only from those people who provide data for a given month. However, the CPS estimates will be biased if the missing observations are not randomly chosen from the population i.e. the dataset is missing not at random (MNAR). In that case, we cannot use maximum likelihood estimation and multiple imputations as they assume that the data is at least missing at random (MAR).

Thirdly, there is a mismatch between the length of a job search that an individual reports in a month and the labor force status that the same individual provides in the previous month. Furthermore, the reported length of unemployment is incompatible with unemployment hazard rates. For instance, from the BLS data, there was a 38 percent probability of an unemployed individual to exit unemployment in the next month in 2011, whereas the probability amongst those unemployed for greater than six months was 31 percent. Hence, although we expect that the mean length of unemployment would be 3 months, the BLS noted that the mean unemployment duration amongst unemployed was 40 weeks or 10 months, starkly diverging from the forecasts based on the hazards reported. Lastly, due to differential preferences, people report certain numbers differently than others. To elaborate, more people on an average state that they have been searching for jobs for 6 months (or 24 weeks), than other people who have been searching for jobs for 23 weeks. Moreover, people are less likely to report an odd number of weeks for unemployed duration than an even number of weeks for the same.

Just as there are inherent flaws in utilizing the unemployment rates as a benchmark for labor utilization, the inflation rate is problematic. Projected values of percent change in CPI is very volatile in the short-run due to transitory shocks such as changes in oil price. Therefore, crossing the inflation threshold is very likely, albeit, unemployment may still be very elevated. As a corollary, a higher inflation target would warrant more favorable unemployment conditions, than those when the inflation threshold is lower. Alternatively, the Fed may decide to communicate to the public why it is not viable to raise the federal funds rate target, notwithstanding the timing of the breach of the inflation threshold. This will make the Fed's actions less credible to the public - as the public had believed that crossing the inflation threshold would stimulate the Fed would tighten the federal funds rate, albeit it didn't. To overcome the issue, the Fed may set a higher threshold for inflation; however, this risks the chance that the Fed will not tighten the policy when inflation is very persistent. In that case, the Fed may decide to base its threshold contingent upon ex-ante inflation levels between one and two years ahead. But this becomes fraught with problems if the Fed's inflation projections substantially deviate from those of the private sector, risking its credibility.

Instead of having a pre-specified unemployment threshold of 6.5 percent, using a range threshold would be apt, "at least as long as the unemployment rate" surpasses the threshold. This implies that the Fed would welcome unemployment levels even when it is below 6.5 percent. Expectations also play a major role - if the Fed announces a policy threshold, but market participants don't materially alter their expectations on knowing the date when the Fed will lift off from its current low-interest rate, then the stimulus would not be very efficacious. This is because if the market participants anticipate protracted periods of the target federal funds rate at zero lower bound and tightening conditions even when the Fed has changed its preferences in the medium run, then stimulus will not have the desired expansionary effect. On the other hand, if FG enables market participants to better understand the FOMC's reaction function, such that they revise their expectations of medium and long term interest rates, aligning with the Fed's future trajectory of short term rates, then the markets will become stable.

Nonetheless, there is a drawback of using the unemployment rate as a measure of slack in the economy. The FRB/US stochastic simulations omit the measurement uncertainty, which is very telling as the natural rate of unemployment is unobserved and is with a degree of uncertainty. If policymakers underestimate the natural rate of unemployment, then they will view that the economy has more slack than is actually, unwittingly overshooting the accommodative monetary policy stance for longer than necessary, spiking inflation. This is germane as policymakers consider the natural rate of unemployment when setting the threshold for the unemployment rate. Particularly, if the inflation expectations are not firmly anchored, then overestimating slack in the labor market will cause measurement error and undermine the FOMC's credibility. Thus, market participants mark up their inflation expectations in the long run.

Forward Guidance through FOMC's statements should explicitly outline in the escape clause that the numbers are conservative thresholds and not triggers i.e. the FOMC may disregard the established thresholds (and continue to keep the federal funds rate $-ffr$ low) if economic and financial conditions are unstable, even if the thresholds are breached. On the flipside, the FOMC may decide to lift off the policy rates even before crossing the threshold when the economy is rebounding faster than expected. The statements should also specify a monetary policy reaction function in place on which $ffr$ will begin to rise from the ZLB. The strongest policy accommodation occurs when a rule includes the inertial behavior than without it.

The Taylor rule is: $R_t^T = r_t^{LR} + \pi_t + 0.5\,(\pi_t - \pi^*) + 0.5\,(y_t - y_t^P)$

The inertial Taylor rule is: $R_t^I = 0.85\,R_{t-1} + 0.15\,[r_t^{LR} + \pi_t + 0.5\,(\pi_t - \pi^*) + 0.5\,(y_t - y_t^P)]$

Here, $R_t^T$ and $R_t^I$ are the nominal federal funds rate characterized by the Taylor rule and inertial Taylor rule, respectively. $r_t^{LR}$ is the long run level of neutral inflation-adjusted federal funds rate, which we expect on average to be consistent with the 2 percent inflation rate and maximum output produced when resources are fully utilized. $\pi_t$ is the 4-quarter moving average price inflation for quarter $t$, and $\pi^*$ is the 2 percent target inflation rate. $y_t$ and $y_t^P$ are the logs of real GDP and potential GDP, respectively, in quarter $t$. $(\pi_t - \pi^*)$ and $(y_t - y_t^P)$ are the inflation and resource utilization gaps, respectively. The advantage of the inertial Taylor rule is that it generates an outcome similar to the one without the inertial behavior, but with a gentler slope - it damps the volatility in the short term $ffr$. Without inertia, the federal funds rate moves immediately without indicating the direction of its movement.

The Taylor rule's intercept is the medium term natural rate of interest $- r_t^{LR} + \pi_t$. Therefore, the Taylor Rule suggests that the *unrate* threshold at which $ffr$ rises above the ZLB hinges on the estimate of the natural rate of interest. Liftoff will happen at a lower *unrate* if the natural interest rate lowers. Forward Guidance will be stimulative if the market participants' expectations alter when the FOMC announces how $ffr$ would change when the thresholds breach. Without changes in expectations of the public, and without clarifying the post-threshold reaction function, the desired stimulative effect of the thresholds will wane.

Currently, the FOMC measures inflation through the rate of change of price index for the total personal consumption expenditure (PCE). However, on a year-to-year basis, the total PCE is extremely volatile. Hence, we should consider alternative methods that reweigh the components of the index to discriminate transitory from persistent movements and lower measured inflation's variance. One variable is the nominal wage growth − *eciwag_rate*, measured through Employee Compensation Index (ECI), and the other is the prime-age employment to population ratio − *epop*. It is a function of *unrate*:

$$epop = labor\ force\ participation\ rate \times (1 - unemployment\ rate)$$

**Establishing the Thresholds of *eciwag_rate* and *epop***

In order to establish an *epop* threshold in the FRB/US model, I had to find a value of *epop* that corresponds to the 6.5 percent level of *unrate*. The analysis encompasses monthly frequencies of the two time series from January 1, 1990 to June 1, 2012. The $corr(epop_t, unrate_t) = -0.95$ indicates a strong negative correlation, which we expect. Before regressing *epop* against *unrate*, I have checked for stationarity of the series as shown in table 1. Here, the null hypothesis is that a series has unit root or is non-stationary.

| Variable | p-value | Inference |
|---|---|---|
| *epop* | 0.7896 | $fail\ to\ reject\ H_0 \Rightarrow non-stationary\ series$ |
| *unrate* | 0.1565 | $fail\ to\ reject\ H_0 \Rightarrow non-stationary\ series$ |
| Δ*epop* | 0.0001 | $reject\ H_0 \Rightarrow stationary\ series$ |
| Δ*unrate* | 0.0082 | $reject\ H_0 \Rightarrow stationary\ series$ |

*Table 1: Test for Stationarity for labor utilization variables*

The large p-values indicate that both *epop* and *unrate* have root, and thus, should be differenced before conducting further analysis. Furthermore, figure 1 indicates that they are graphically autocorrelated as the spikes gradually diminish within the significance bounds due to trends in *epop* and *unrate*.

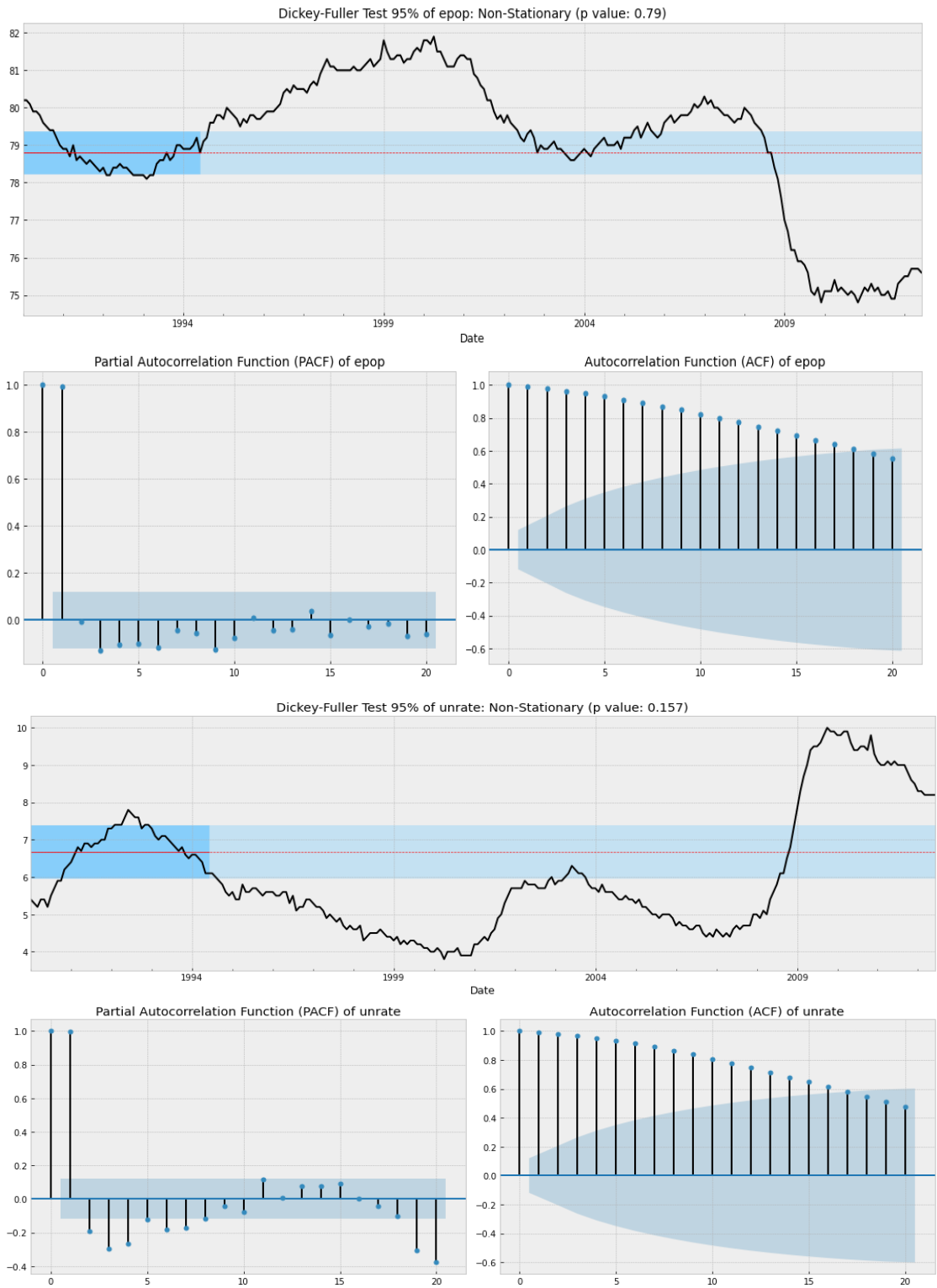

*Figure 1: ACF and PACF plots of epop (above) and epop (below)*

Each of these variables don't have constant mean and constant variance i.e they don't exhibit mean reversion and have stochastic trends. Figure 2 shows the graph of the differenced *unrate* and *epop,* also called *unrate_diff* (or $\Delta unrate$) and *epop_diff* (or $\Delta epop$) respectively, which also indicate the absence of unit root.

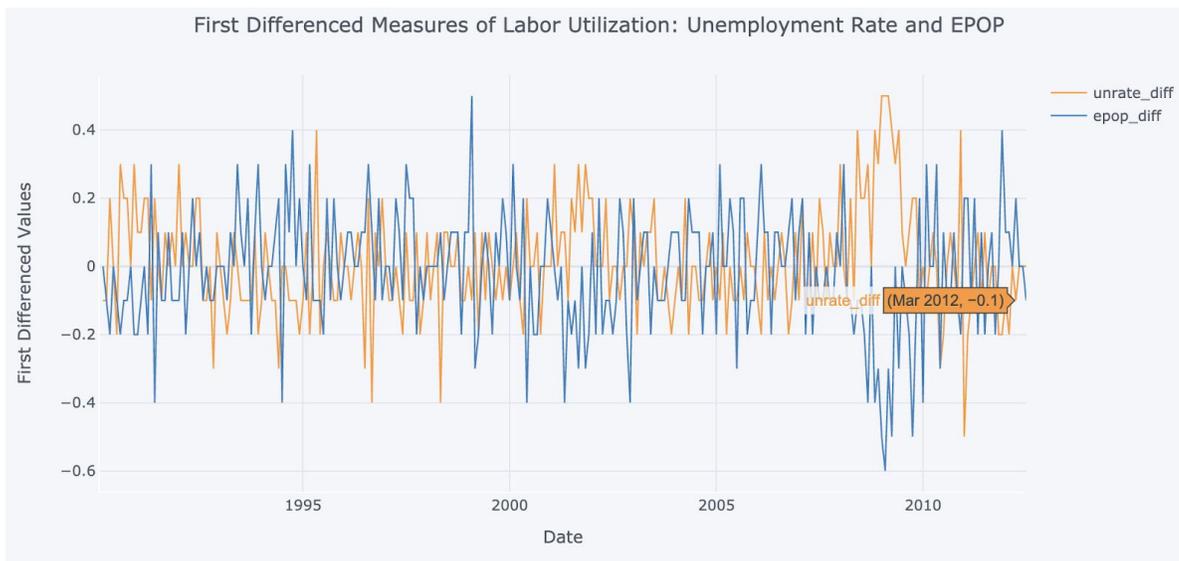

*Figure 2: Time series of the first differenced unrate and epop are stationary*

I have regressed $\Delta epop$ onto $\Delta unrate$ to generate the comparable value of *epop* for an *unrate* of 6.5 percent i.e. I will estimate $epop_{thresh}$ for a given level of $unrate_{thresh}$.

$$(M1) \quad \Delta epop = -0.0116 - 0.5329\, \Delta unrate \qquad Adjusted\ R^2 = 0.225$$
$$\phantom{(M1)\quad\quad} (0.009) \quad\ (0.061)$$

$\Delta unrate$ is significant at the 0.05 significance level. I have rewritten the equation to incorporate the *unrate* threshold:

$$epop_t - epop_{thresh} = -0.0116 - 0.5329\,(unrate_t - unrate_{thresh})$$

Inputting all the values of $epop_t$ and $unrate_t$ from the series will yield a range of values of $epop_{thresh}$. For example, when $epop_{1990/01} = 80.2$ and $unrate_{1990/01} = 5.4$, and $unrate_{thresh} = 6.5$, then:

$$80.2 - epop_{thresh} = -0.0116 - 0.5329\,(5.4 - 6.5) \Rightarrow epop_{thresh} = 79.62541.$$

The descriptive statistics of $epop_{thresh}$ shows that an ideal range of $epop_{thresh} = (78.39,\ 81.07)$. An alternative way to estimate $epop_{thresh}$ would be via a leveled regression of *epop* onto *unrate*. Generally, OLS on variables with stochastic trend will yield misleading (spurious regression) results as OLS relies on the assumption that observations are *i.i.d,* which is not the case with *epop* and *unrate*. Furthermore, the t-statistic of these regressors are no longer normally distributed, even for large samples. As a corollary, standard hypothesis tests are no longer valid for non-normal distributions. However, the cointegration tests below show that we can regress the levelled values of *epop* on *lur* without introducing biases and inferential problems as both the variables share a common stochastic trend. Cointegration analysis enables us to examine the long-run linkages among the economic variables, considering the adjustments in the short-run to deviations from the equilibrium in the long run. So, I have performed the Johansen cointegration test to establish a long run relationship. This implies that we can assume a long run relationship in the model even if the series are drifting apart or trending either upwards or downwards.

In the Johansen Cointegration Test:

$H_0$ : series are not cointegrated i.e there is no cointegration equation

$H_A$ : series are cointegrated

We reject the null hypothesis if the value of the trace and maximum eigenvalue statistics are greater than the 0.05 critical value statistics. From the unrestricted cointegration rank test, $r = 0$, implies that there are no hypothesized cointegrating relations. The trace statistic of $16.9452 > 16.1619 -$ critical value statistic. Thus, we reject the null hypothesis that there are no cointegrating relations. When $r = 1$, there is at most one hypothesized cointegrating relation. The trace test statistic of $0.9839 < 6.6349 -$ critical value statistics, fails to reject the null hypothesis. This signifies that there is at most one cointegrating relation between *epop* and *unrate*. The results from the maximum eigenvalue test also confirms the aforementioned conclusion.

Because *unrate* and *epop* are cointegrated, they exhibit a long run relationship and can be linearly combined. There exists a linear combination of *epop* and *unrate* that is stationary. While shocks in the run may affect the movement in the individual series, they would converge with time i.e. they get back to some form of equilibrium in the long run. Albeit both *unrate* and *epop* are $I(1)$, $epop - \beta$ *unrate* is $I(0)$, enabling us to regress one against another, without generating spurious regression results.

$(M2) \quad epop = 85.6303 - 1.0896 \text{ } unrate \qquad \text{Adjusted } R^2 = 0.916$

$\qquad\qquad\quad (0.126) \quad\ (0.020)$

The coefficients are significant at the 0.05 critical value. Thus, $epop_{thesh} = 78.5479$ when $unrate_{thresh} = 6.5$.

I have also estimated the thresholds of inflation measures. Currently, the FOMC measures inflation through the rate of change of chain-type price index for the total personal consumption expenditure (PCE), known as *pce_rate* in this analysis. However, on a year-to-year basis, the total PCE is extremely volatile. Hence, we should consider alternative methods that reweigh the components of the index to discriminate transitory from persistent movements and lower measured inflation's variance. One variable is the nominal wage growth - *eciwag_rate,* measured through Employee Compensation Index (ECI). As a better rule-of-thumb forecast of headline inflation, it excludes a fixed percentage of severe price changes.

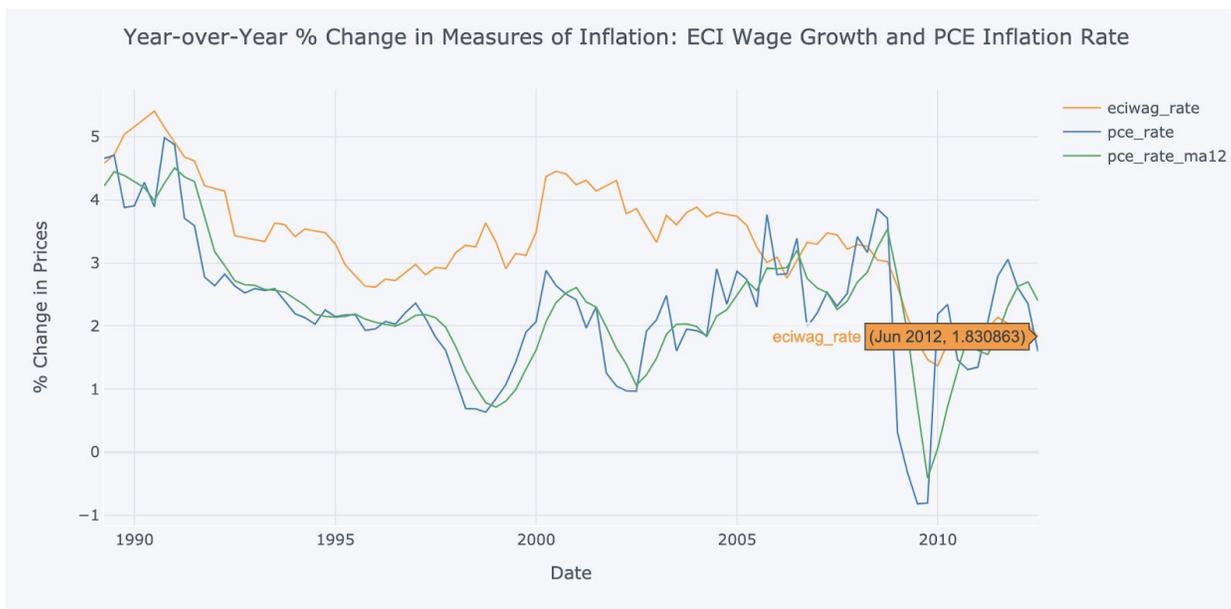

*Figure 3: Volatile measures of inflation rate relative to wage growth*

Although the 4 quarter (or 12 months) moving average of PCE — $pce\_rate\_ma12$, smooths the series and tempers the volatility arising from the supply shocks such as fluctuations in food, oil and other commodity prices, it is still more volatile than the ECI wage rate. The graph in figure 3 above compares the percentage change in 4 quarter moving average PCE: chain-type price index (index 2012 = 100) with the percent change in PCE rate without the moving average. The latter is highly volatile as depicted by strong fluctuations in its trend and standard deviation in the descriptive statistics shown below:

|  | pce_rate | eciwag_rate | pce_rate_ma12 | unrate | epop |
|---|---|---|---|---|---|
| Mean | 2.380 | 3.418 | 2.402 | 79.054 | 6.033 |
| Standard Deviation | 1.096 | 0.878 | 0.971 | 1.842 | 1.618 |
| Minimum | −0.817 | 1.368 | −0.411 | 74.8 | 3.8 |
| 25% | 1.920 | 2.972 | 1.982 | 78.6 | 4.7 |
| 50% | 2.328 | 3.387 | 2.320 | 79.4 | 5.6 |
| 75% | 2.858 | 3.878 | 2.829 | 80.1 | 6.9 |
| Maximum | 4.958 | 5.405 | 4.506 | 81.9 | 10 |

*Table 2: Summary statistics of inflation and labor utilization variables*

The variables that measure inflation — $pce\_rate$ and $eciwag\_rate$, are very autocorrelated as the spikes (figure not shown) in the correlogram cross the significance level for five lags, implying that the serial correlations amongst the lags are different from 0. Thus, the variables are non-stationary and I have differenced them to remove the unit roots.

| Variable | p-value | Inference |
|---|---|---|
| $pce\_rate$ | 0.3767 | $fail\ to\ reject\ H_0 \Rightarrow non-stationary\ series$ |
| $eciwag\_rate$ | 0.4902 | $fail\ to\ reject\ H_0 \Rightarrow non-stationary\ series$ |
| $\Delta pce\_rate$ | 0.0101 | $reject\ H_0 \Rightarrow stationary\ series$ |
| $\Delta eciwag\_rate$ | 0.0006 | $reject\ H_0 \Rightarrow stationary\ series$ |

*Table 3: Test for Stationarity for inflation rate and wage growth variables*

The non-spurious difference regression of $\Delta eciwag\_rate$ against $\Delta pce\_rate$ is:

$$(M3) \quad \Delta eciwag\_rate = -0.0365 + 0.0624\ \Delta pce\_rate \qquad Adjusted\ R^2 = 0.23$$
$$\quad\quad\quad\quad\quad\quad (0.009)\quad\quad (0.061)$$

To check if the regression model $M3$ satisfies the OLS assumptions and that the estimators are the best linear and, I've constructed the model's diagnostics plots. Firstly, the null hypothesis of the Harvey - Collier Multiplier test for linearity tests if the regression is linear. From the $p-value = 0.178 > 0.05$, we fail to reject the null hypothesis and conclude that $M3$ is linear. The randomness in the residual plot (not shown) further corroborates the presence of linearity. Secondly, the normal probability plot (not shown) assesses if the residuals visually depart from normality and the model's good fit implies that the normality assumption is upheld.

Lastly, the White's test for heteroskedasticity tests if the variance of the residuals are constant or if they change over time. Both the LM - test and F-test p-values of 0.69 and 0.7, respectively are greater than the 0.05 significance level, implying the residuals are homoscedastic. The expected value of the residuals, $E(\varepsilon) = 5.2867 \times 10^{-19}$, which is very small and approaches 0. Thus, the model satisfies the Gauss Markov Theorem and we can use $M2$ to estimate $eciwag\_rate_{thresh}$. Rewriting the equation to incorporate $pce\_rate_{thresh} = 2.5$ to estimate $eciwag\_rate_{thresh}$ results in the following equation:

$$eciwag\_rate_t - eciwag\_rate_{thresh} = -0.0365 + 0.0624\ (pce\_rate_t - 2.5)$$

From the descriptive statistics of $eciwag\_rate_{thresh}$, an appropriate range of $eciwag\_rate_{thresh} = (3.07,\ 4.24)$.

|  | $epop_{thresh}$ | $eciwag\_rate_{thresh}$ |
| --- | --- | --- |
| Mean | 79.32 | 3.61 |
| Standard Deviation | 2.67 | 1.07 |
| Minimum | 72.99 | 1.26 |
| 25% | 78.39 | 3.07 |
| 50% | 79.96 | 3.47 |
| 75% | 81.07 | 4.24 |
| Maximum | 83.35 | 6.53 |

*Table 4: Descriptive statistics of epop and eciwag_rate thresholds*

I did not construct a leveled regression model because *pce_rate* and *eciwag_rate* are not cointegrated. When $r = 0$, then the trace statistic of $23.1932 < 18.3985 -$ critical value at 0.05 significance level. When $r = 1$, the trace statistic of $7.0524 < 6.63949$. So, we reject the null hypothesis that there is at most one cointegrating relationship. This signifies that *pce_rate* and *eciwag_rate* are not related in the long run and the series diverge from one another.

Moreover, widespread misclassification errors in the CPS data have yielded imprecise labor force participation status of the survey participants each month (Feng and Hu 2013). The largest errors arise when participants in the survey have been misclassified from unemployed to employed and from unemployed to not in the labor force. It is particularly hard to classify marginally attached workers, such as part-time and discouraged workers. After correcting for errors, they show that the official unemployment rate substantially underestimates the true *unrate* on average by 2.1 percent from January 1996 to August 2011. Furthermore, *unrate* is very sensitive to the business cycles as higher *unrate* during a downturn underestimate the true level of *unrate* by a greater magnitude than in expansionary phases of the business cycle.

Establishing a numerical threshold of *unrate* is contingent upon comprehensively understanding the labor force participation rate - *lfpr*. Forecasting *lfpr* is hard as it is difficult to distinguish between the change due to cyclical and trend components. I have decomposed *lfpr* as an additive model where:

$lfpr_t = level_t + trend_t + seasonality_t + noise_t$

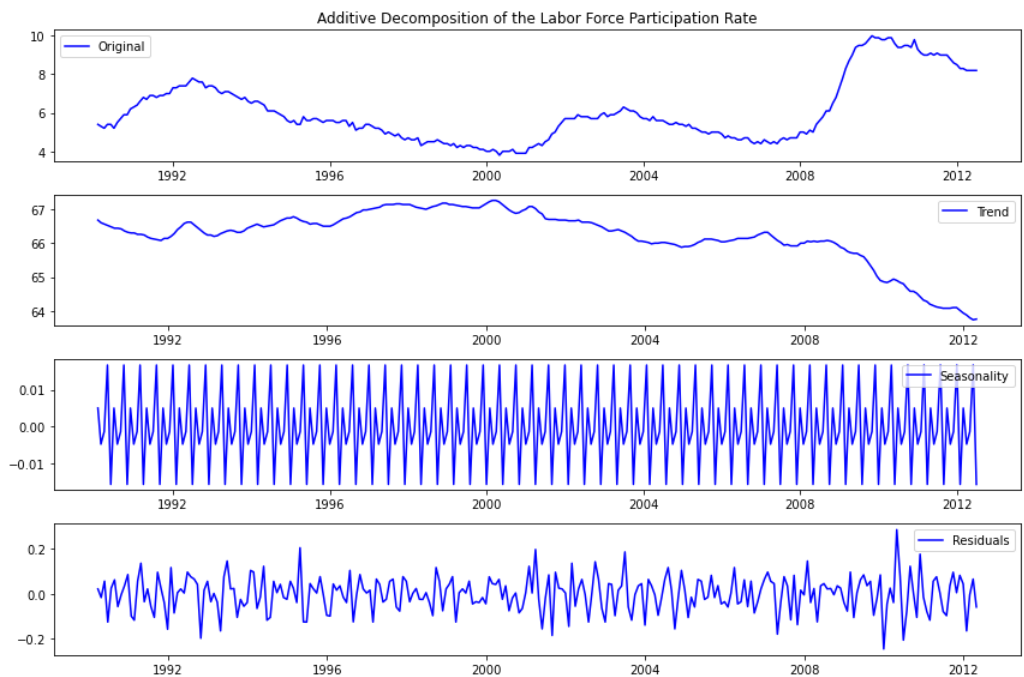

*Figure 4: Additive decomposition of lfpr*

Monetary policy tools will be efficacious to the point that the patterns in *lfpr* are seasonal (or cyclical), whereas fiscal policy would be more appropriate to accommodate the structural patterns affecting *lfpr*. As a corollary, *lfpr* might spike when *unrate* declines. Alternatively, *unrate* might reduce albeit minimal growth rate of employment as more people flow from unemployment to non-participation (or not in the labor force). Like *lfpr*, the *unrate* is a noisy indicator of labor market conditions. Policymakers assign *unrate* threshold by gauging the non-accelerating inflation rate of unemployment (NAIRU), also known as the natural rate of unemployment. Mounting evidence ascribes to its decline, consequently, diminishing the threshold for *unrate*. Furthermore, NAIRU estimates are imprecise and not robust (Staiger, Stock, and Watson 1997, 2001; Stock 2001). We acquire NAIRU estimates from those of the Phillips Curve, whose inverse relationship between the unemployment gap $-unrate\_gap$ and *pce_rate* is flattening over time. The OLS regression results showcase a weak relationship between *unrate_gap* and the growth rate of PCE - *pce_rate*. Both $M4$ and $M5$ represent variations of the Phillips curve.

$$(M4) \quad pce\_rate = 2.5271 - 0.013 \, unrate\_gap \qquad Adjusted \, R^2 = 0.084$$
$$(1.929) \quad (0.304)$$

$$(M5) \quad eciwag\_rate = 3.6518 - 0.0208 \, unrate\_gap \qquad Adjusted \, R^2 = 0.362$$
$$(0.074) \quad (-0.0208)$$

Furthermore, the low correlation between *unrate* and *pce_rate* of $-0.213$ signifies that policymakers should not emphasize strongly on thresholds of *unrate* and *pce_rate*. From the correlation matrix, the association between

*eciwag_rate* and unemployment variables is stronger than that with *pce_rate,* signifying that *eciwag_rate* is a better variable than *pce_rate*. For instance:

$corr(unrate, eciwag\_rate) = -0.547 > corr(unrate, pce\_rate) = -0.213$ and

$corr(unrate\_gap, eciwag\_rate) = -0.606 > corr(unrate\_gap, pce\_rate) = -0.213$

| Variable | unrate_gap | pce_rate | unrate | eciwag_rate |
|---|---|---|---|---|
| unrate_gap | 1.000 | −0.304 | 0.975 | −0.606 |
| pce_rate | −0.304 | 1.000 | −0.213 | 0.568 |
| unrate | 0.975 | −0.213 | 1.000 | −0.547 |
| eciwag_rate | −0.606 | 0.568 | −0.547 | 1.000 |

Table 4: Correlation matrix

Moreover, upon decomposing *pce_rate*, we see a considerable proportion of short-run noise (residual) in the series, making it an unreliable time series on which to base the thresholds.

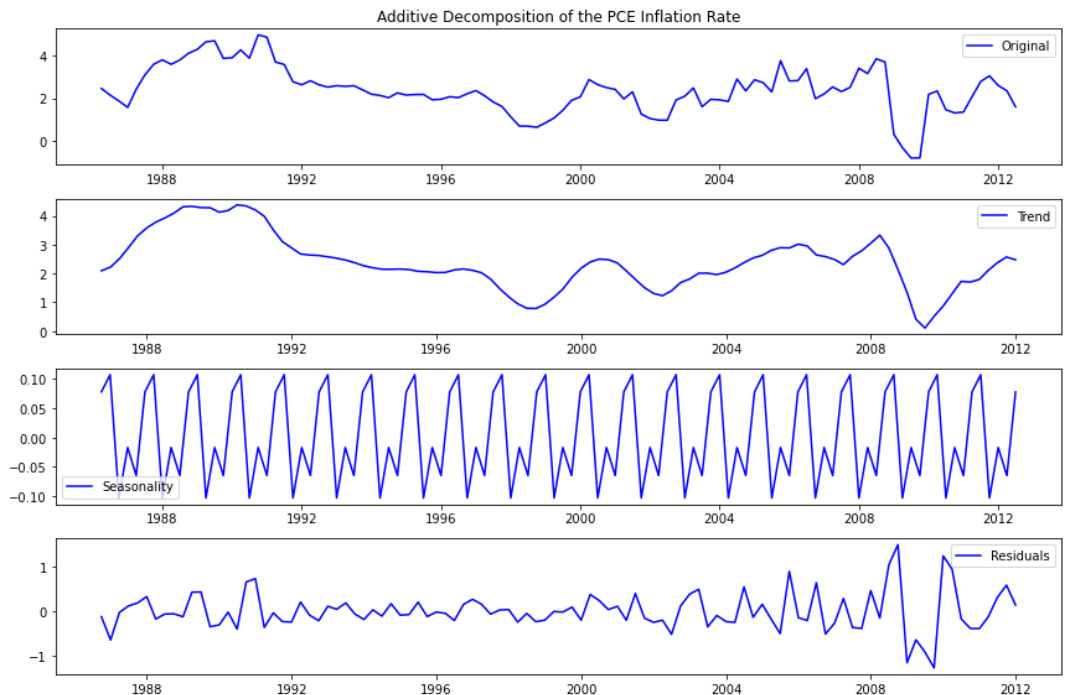

Figure 5: Additive Decomposition of *pce_rate*

Preliminary estimates on economic variables such as GDP, *unrate, lfpr,* etc measure aggregate resources are often subject to revision later once statistical agencies incorporate complete information in published data. Economists extract supply-side estimates of NAIRU, total factor productivity, and potential real GDP from these evolving sources of data.

Consequently, the output gap, which is the difference between the actual and potential GDP as a percentage of potential GDP, is often revised by large amounts due to the inherent difficulty in forecasting potential GDP. Theoretically, large positive output gaps indicate the economy is exceeding its maximum potential level of output, spiking inflationary pressures. With imprecise measures of the output gap, policymakers may fear expected stronger inflation than in actuality and wrongly raise the federal funds rate. Hence, we need to consider a wide array of indicators in assessing the labor market conditions and incorporate their thresholds to determine the timing of the liftoff of *rff* from the ZLB.

Subsequently, I've discussed the mechanism behind the FRB/US model and investigated the effects of incorporating $eciwag\_rate_{thresh}$ in the FRB/US model, and compared the trajectory of macroeconomic variables with the output from the original threshold of *pce_rate*.

**A Brief Introduction to the FRB/US Model**

FRB/US is a New Keynesian structural policy model that has roughly 375 equations, including 50 "core" stochastic equations. The equations and identities model the variables in sectors such as the labor market, financial sector, foreign activity, business investment, fiscal and monetary policies, etc. The Federal Reserve Board uses it to computationally simulate the impact of "what-if" fiscal, monetary and global shocks and project their outcomes on macroeconomic variables, given certain assumptions. This paper shows the ex-ante effects of a few shocks under a wide array of alternative policy responses. Firms and households in the model are forward-looking as they formulate their ex-ante decisions on sales, income, prices, and financial conditions based on their future expectations of the economy. Instead of reacting instantaneously, they respond gradually as costs of capital installations, contracts, etc create frictions that stagnate the process. Thus, markets don't fully utilize the capital resources and labor as they fail to quickly clear when shocks disturb the economy's equilibrium.

Expectations play a predominant role in affecting prices in the financial markets, although the model assumes that the financial decisions are insensitive to frictions, due to minuscule transactional costs. Expectations can either have a forward-looking model consistent (MCE) solution or a VAR solution to the optimization problems. MC expectations are rational wherein the firms and households understand the functioning of the economy very well, and expectations are equal to the projections generated by the FRB/US model. Alternatively, expectations can be based on a partial understanding of the economy, depicted by a small forecasting model that consists of a few essential macroeconomic variables and their lagged values. Resembling vector autoregression (VAR), hence these expectations are known as VAR expectations. Herein, the expectations of firms and households are contingent upon their knowledge of how the three variables interact – inflation rate, federal funds rate, and cyclical economic state. These expectations have a backward-looking representation as follows:

$$Y_t = \sum_{i=n_o}^{n_1} \beta_i X_{t-i}$$

$X_t$ = vector of variables, $\beta_i$ = coefficient matrices, $Y_t$ = expectation of a variable.

In the FRB/US model, inflation, which is derived from a variant of the New Keynesian Phillips curve, is very slow to adjust. It is based on the historical and projected inflation, alongside the expected resource utilization and the markup of prices over trend costs of labor. The firms strive to pay the wages that equate to the marginal product and price their final goods and services as a markup over the trend unit input costs. Yet, labor costs, among other factors, are harbingers as they create the friction that curtails the speed at which prices and wages adjust to shifts in supply and demand. FRB/US equations incorporate this "sticky-price" nature to govern the inflation rate's reaction when economic conditions change. Consequently, changes in the short-term nominal interest rates temporarily influence the real interest rates, affecting real prices and yields on various securities, which indirectly impact economic growth. The short-run frictions hinder the labor market to be in equilibrium. In that case, wage and price inflation are susceptible to increase when *unrate < nairu,* and to reduce when *unrate > nairu*. However, other factors such as movements in energy and commodity prices, supply shocks may alter wage and price inflation changes.

The FRB/US simulations that I have constructed consists of thresholds of *eciwag_rate* only, and not *epop*. In order to establish thresholds of *eciwag_rate* In the model, I had to alter the existing equations and add a few new variables. The variables that affect the threshold-based policies are *dmptunrate* and *dmptpi* – the monetary policy indicators of unemployment and PCE inflation rates, respectively. I replaced *dmptpi* with *dmpteci* – the monetary policy indicators of the ECI wage rates.

$$dmpteci = 1\{eciwag\_rate > eciwag\_rate_{thresh}\}\ {}^2$$
$$dmptunrate = 1\{unrate > unrate_{thresh}\}$$

These alterations will also affect the policy rule for calculating the federal funds rate, the various monetary policy reaction functions such as the Taylor and the inertial Taylor rule.

Adding these variables will affect *dmptrsh* - switch variables for monetary policy.
$$dmptrsh = 1\{\exists\ :\ a\ threshold\}$$
So, *dmpteci* =1 delays the liftoff of the federal funds rate from the ZLB until either *unrate* falls below the critical rate – *unrate_{thresh}*, or *eciwag_rate* rises above the threshold – *eciwag_rate_{thresh}*. Other endogenous trigger variables that are affected are:

$$dmptmax = max\ (dmptunrate,\ dmpteci).$$
Thus, $dmptmax = 1$, when either $unrate_{thresh}$ or $eciwag\_rate_{thresh}$ is breached.

---

[2] I acquired the data on the ECI wage growth (as forecasted in 2012) from the Greenbooks Datasets.

$$dmptr = max(dmptmax, dmpteci_{t-1}).$$

So, *dmptr* is initially 0 and remains at that value until either one of the thresholds is crossed, after which it equals to 1.

**Results of the FRB/US Simulations with *eciwag_rate* and *pce_rate* Thresholds**

In all the graphs, I have simulated[3] and compared the impact of different macroeconomic shocks on the path of the VAR expectations from 2012Q3 to 2017Q4, keeping the unemployment rate threshold at 6.5 percent. Each of the graphs, that show the impulse response functions, has three trajectories:
1. Consensus baseline: which is the Summary of Economic Projections (SEP) forecasts;
2. Projected PCE inflation rate threshold − $pce\_rate_{thresh} = 2.5\%$, which is currently adopted under Evan's Rule
3. ECI wage rate threshold − $eciwag\_rate_{thresh} = 3.5\%$.

The ECI wage rate is the annualized growth rate of hourly compensation, measured through the employment cost index, whose equation is implicitly derived from the model of bargaining over the real wage. Based on implicit and explicit multi-period contracts, wages are less flexible than prices − they gradually adjust to their equilibrium level. In periods of high *unrate,* workers' bargaining power recedes relative to periods of low *unrate,* diminishing their ability to demand higher wages.

The first simulation measures the impact of a strong negative aggregate demand shock. In both $pce\_rate_{thresh} = 2.5$, and $eciwag\_rate_{thresh} = 3.5$, I have assumed that the Taylor rule monetary policy reaction function − *rfftay*, prescribes the path of the federal funds rate. In the former case, I have activated the fiscal policy that adjusts the tax rates such that the ratio of the federal government's surplus to GDP gradually stabilizes at a specified value − *dfpsrp* = 1. In the latter case, I have activated the fiscal policy that makes no adjustments and is suitable in the short-run only − *dfpex* = 1. In both cases, I have inputted the values of the shock derived from the residuals of the equations for 2008Q4 to 2009Q3 that are given in the FRB/US model itself.

In the first case, the federal funds rate lifted off from the ZLB in ~2016Q2, which is a delay of approximately two quarters from the baseline case, where the FOMC lifted off the policy rates in 2015Q3. In the third case, $eciwag\_rate_{thresh} = 3.5\%$, the delay is further prolonged to 2017Q1. As opposed to the path in the Evan's Rule, wherein *ffr* steeply rises, *ffr* gradually increases. The year to year percent change in the real GDP plummets due to the strong negative AD shock in 2012-2013, before reaching levels comparable to those in the baseline. The shock also increases the unemployment rate − *unrate,* from 8 to 9.6 percent in its peak in 2013Q1; and consequently diminishes the employment to population ratio $-epop$ to 0.572. The 10-year treasury yield $-rg10$, PCE inflation rate $-pce\_rate$, core PCE rate $-corepce\_rate$, and the ECI wage rate $-eciwag\_rate$, also remain below their baseline forecasts, which is consistent with the usual macroeconomic theory. Lower AD generates lower output as the consumers' demand for goods and services diminishes. As a corollary, firms don't have to hire new workers, and could possibly fire a few of their existing workers, raising the unemployment rate. Moreover, falling aggregate demand may contract the demand for the US treasuries of varying maturities, including the 10-year Treasury. A depressed demand may diminish the price of the security, raising $rg10$ as price and yields are inversely correlated. Since the rise in $rg10$ affects long-term fixed rates such as the 10-year mortgage rate, housing becomes less affordable for consumers. Thus further diminishes the consumer demand which composes of ~70 percent of the US gross domestic product.

---

[3] I constructed impulse response function graphs from FRB/US simulations in EViews, and the results are in the Appendix.

The second set of graphs calibrate the impact of a transient $20/barrel increase in oil prices. In both the scenarios, I have assumed that the federal funds rate adheres to the Taylor rule after crossing $eciwag\_rate_{thresh}$ and $pce\_rate_{thresh}$. Setting $dfpdbt = 1$, I have set the fiscal policy to adjust the tax trends in a way that stabilizes the ratio of government debt to GDP in both the cases. The graphs depict a decline in the real GDP by ~1.8 percent from the peak as reached in the consensus baseline, but $rgdpch$ declines more under Evans Rule 2.0 than in the currently established rule. Unlike $pce\_rate$, $corepce\_rate$ doesn't fluctuate very much and is less volatile as it excludes the price of food and energy, including oil. Rising oil prices feed directly into upsurging prices of gasoline oil, diesel, and other forms of energy. These raise the utilities costs for both consumers and producers, denting the profit margins. Thus, workers demand higher wages, but because wages are sticky and adjust more slowly than prices, real wages fall, depressing the consumer demand for goods. As consumption constitutes 70 percent of the GDP, its decline also reduces $rgdpch$, which, from the graph, is stronger when $eciwag\_rate_{thresh} = 3.5\%$. $pce\_rate$ under the wage rate threshold is about 0.4 percent higher from the baseline when oil prices in 2012, before plunging and converging at 1.45 percent in 2017-Q4. $ffr$ lifts from the ZLB much sooner in 2014-Q1 than that in the baseline case, while $ffr$ lifts in 2014Q3 when $pce\_rate_{thresh} = 2.5\%$ not only because of mounting current inflationary trends, but also because of rising inflation expectations. As the transitory weakness in the aggregate spending propels the Fed to stabilize the disequilibrium, it starts to lower the pace of the hike after 2016-Q3 under $pce\_rate_{thresh} = 2.5\%$, while it lessens the policy rate from 3.25 to 2.75 percent in 2017-Q4 under $eciwag\_rate_{thresh} = 3.5\%$. Seeing that both the inflation rates become close to the baseline after one year, the decline in the $ffr$ is meant to offset the contractionary disturbances due to the spike in oil prices.

Another negative supply-side shock is a lower labor force participation rate $-lfpr$. As in the case of the negative oil shock, $ffr$ lifts early although the slope is less steep. Again, the federal funds rate follows the prescriptions of the Taylor rule after the threshold is breached. Whilst $unrate$ in the current PCE inflation threshold case appears to move parallely with $unrate$ from the consensus baseline case, the respective $epop$ paths tend to diverge. $unrate$ is lower from the baseline for two years from 2012Q2 until 2014Q2 from ~8.2 to 6 percent in both the thresholds. From 2014Q3, the decline in $unrate$ diminishes when the economy follows the ECI wage rate threshold. In both the thresholds, $eciwag\_rate$ is generally higher and $pce\_rate$ is lower. Historically, lower $unrate$ is associated with ascending wages as a deficit of workers in the firms cause firms to raise wages to attract qualified workers and satisfy the consumer demand, typically evident in a tight labor market. However, quite bizarrely, as also observed in the oil shock, the forecasts of *rg10* fluctuate (more than usual) like waves from 2015Q4 to 2016Q4.

Next, I have quantified the effects of unanchored inflation expectations under the inertial Taylor rule − *rffintay*, when $pce\_rate_{thresh} = 2.5\%$, and the Taylor rule without the inertial behavior when $eciwag\_rate_{thresh} = 3.5\%$ I have changed the 10-years expected PCE price inflation (from the Survey of Professional Forecasters) − *ptr*, to deviate from its current level by 5 percent; implying that the long-run inflation rate has significantly responded to new information and market surprises. If the inflation expectations were anchored, then $\Delta ptr = 0$ over time. Due to the inertial nature of the policy reaction function, *ffr* stays in the ZLB for a quarter longer than in the original Evan's Rule, than the proposed rule. Here, the liftoff occurs in 2014Q1 and shoots to 4 percent in 2017Q4 which contrasts with the slow rise in *ffr* to 2.8 percent when the inertial behavior guides the rates. Just as in all the previous simulations, the percent change in the real GDP under both thresholds is still lower than those in the consensus baseline scenario.

The fifth simulation assesses the influence of the Taylor rule with the unemployment gap monetary policy reaction function − *rfftlr*, after both the thresholds are breached, keeping other model specifications constant.

The Taylor rule with unemployment gap is: $R_t^T = r_t^{LR} + \pi_t + 0.375\,\pi_t - 0.5\,\pi^* + 1.1\,(unrate_t - nairu_t)$

This version of Taylor rule calculates the federal funds rate by adding the equilibrium real funds rate with the four-quarter moving average of ex-post inflation rate. Then, we adjust this value since the actual inflation rate − $\pi_t$ deviates from the target rate − $\pi^*$, and the unemployment rate deviates from the natural rate of unemployment − *nairu*. Here, *ffr* rise in 2015Q4 in all three cases. *rgdpch* is lower in both the threshold conditions than in the baseline until 2016-Q2 Albeit the real GDP growth rate is maximum for up to two quarters from 2015-Q2 at 3.6 percent, the timing and of the peak is slightly variant in either of the threshold conditions. Under both the thresholds, *rgdpch* reaches the apex at 3.5 percent in 2017-Q1 and then tails off. *eciwag_rate* and *epop* rise rapidly under both the current and proposed Evan's Rule. Also, there is a minuscule difference in the trajectory of the inflation variables in both thresholds, presumably because they lift from the ZLB around the same time. There is a burgeoning gap in the paths of the 10-year treasury yields - the yields from both the thresholds are relatively low than those from the baseline, and the rate (when PCE threshold is 2.5%) is slightly lower than when $eciwag\_rate_{thresh} = 3.5\%$ . It appears that the demand for long term treasuries had considerably risen post-2012, as that drives the price up, lowering the yields. The unemployment rates from both the thresholds are much lower, reaching ~4% in 2017-Q4, as opposed to the baseline level of ~5.5% in the same time. Consequently, the *epop* is higher under both the simulations.

The sixth simulation is based on the forward-looking model consistent expectations (MCE). Here, the coefficient values are based on the discounting weights. I have considered that all asset pricing, wages, and price equations have MC solutions. I modeled the impact of a 50 bps hike in the *ffr* wherein the FOMC adheres to the Taylor Rule and inertial Taylor rule under *pce_rate*$_{thresh}$ = 2.5% only. This lifts the *ffr* from the ZLB in 2012-Q2, but under the Taylor rule only, *ffr* soars to 1.8 percent in 2013-Q3, before plunging downwards in a parabolic trajectory to -40 pbs. This (unattainable) monetary policy is a reaction to offset the contractionary economy as a result of the premature hike − lower *rgdpch*, higher *unrate,* depressed inflation rates and soaring *rg10*. Although, *ffr* under the inertial behavior rises relatively to 0.62 percent, after which it falls to -0.1 percent in 2015-Q4 and then rises. This premature hike in 2012 diminishes *rgdpch* considerably, raising *unrate* at the apex of 8.45 percent under the Taylor rule. The worsened job market shrinks the purchasing power of consumers, who now have lower disposable income, lowering the inflation rates and *eciwag_rate*. For example *pce_rate* dramatically fell from 1.75 percent in 2013-Q1 to 0.8 percent in 2017-Q4. In all these graphs, the macroeconomic variables without the inertial behavior deviate considerably than seen in the inertial behavior and consensus baseline.

Then, I modeled the simulated paths of the variables under the optimal control policy and contrasted the results from the SEP-consistent baseline forecasts. OC Policy consists of solving a large scale macroeconomic model to calculate the trajectory of the federal funds rate that minimizes the inflation and unemployment rate deviations from their respective targets. In this approach, the Fed uses choses the level of federal funds rate such that the unemployment and inflation rate targets are fulfilled. Policymakers keep the interest rates low as long as the unemployment rate is farther away from its target than inflation is, even if this causes the inflation rate to overshoot its 2 percent objective for some time.

The seventh simulation checks how the Summary of Economic Projections (SEP) baseline forecast would change if the FOMC selects the trajectory of the federal funds rate set by the optimal control (OC) method to minimize a quadratic loss function. The loss function's role is to penalize equally weighted squared deviations of the inflation rate from its 2 percent target, squared deviations of the unemployment rate from its natural rate, and squared quarterly changes in the federal funds rate.

$$L_O = \sum_{t=0}^{T} 0.99^t \left( w_\pi (\pi_t - 2)^2 + w_u (u_t - u^*)^2 + w_r (\Delta r_t)^2 \right)$$

Agents following the model consistent (MC) expectations in the SEP baseline assume that the federal funds rate initially adheres to a baseline path; thus they accordingly anchor their baseline expectations. When the OC simulation starts, the agents revise their expectations such that they are consistent with the path that the federal funds rate takes under optimal control. Therefore, the policy actions announced are fully credible and agents have rational expectations. Running the simulation with ZLB, I added a penalty term to the loss function and assumed unequal weights on the three terms of the loss function in the OC policy simulation − $w_\pi = w_u = 1,$ and $w_r = 10.$ Under the OC policy, *ffr* rises from the ZLB in

2016Q1, a quarter away from the liftoff proposed under the consensus baseline scenario. A later liftoff timing contributes to an increase in the GDP growth rate relative to the baseline case by ~0.4 percent. Consequently, *unrate* diminishes faster and reaches at 4.7 percent in 2017Q4 in the former, while it is still 5.5 percent in the latter. A lower *unrate* and higher *rgdpch* creates inflationary pressures, spiraling both the core PCE and overall PCE inflation rates by ~0.4 percent. The ECI wage rate robustly rises by ~0.6 percent when compared to the ECI wage rate in the baseline scenario, most likely because a tight labor market incentivizes workers to raise hourly compensation to attract more workers. Finally, *rg10* is lower than its movement in the consensus baseline simulation by ~0.4 percent. We can attribute this to the rising demand of treasuries which raises the prices, thereby, pushing the yields down. As a feedback mechanism, lower yields drive down the long term mortgages and other commercial rates, enticing consumers to refinance their existing loans/mortgages and boost the aggregate demand.

The OC policy doesn't consider an overly-accommodative policy which may be a harbinger for asset bubbles. If the Fed projects that the economy will function with full capacity by a certain year, then guiding the federal funds rate towards zero might wrongly signal the investors to indulge in excessive risks that may dislocate long-term inflation expectations. Consequently, the Fed will have to aggressively tighten its policy. In this case, the Fed's policy of keeping the rates at ZLB would be stretched to a limit and a method to depress the short end of the curve could dislocate the longer end. Whereas optimal control simulations are informative, they hinge on a specific type and features of the model chosen, and a range of simplistic yet possibly unrealistic assumptions. So, optimal control policies may not be robust to misspecifications in the model as they are constructed (Williams and Orphanides, 2008). It also ignores uncertainty about the model's specification. Thus, it would be imprudent to place too much weight on these factors.

OC policy works well when the public's perceptions match those indicated by rational expectations. However, assuming rational expectations, optimal control policy performs poorly in a model wherein the public has imperfect information about the state of the economy, and therefore, must learn. Then, their expectations may deviate from those implied by rational expectations, causing the finely - tuned optimal control policy to go astray. Furthermore, if we implicitly assume that inflation expectations will always be well-anchored, then optimal control policy reacts inadequately when inflation is very volatile. Resultantly, the inflation rate becomes excessively persistent and deviates considerably from its target. William and Orphanides (2008) found that we can make optimal control methods more robust to learning by emphasizing less on the stability of real output and interest rate relative to inflation in the loss function. If the bias present in the weights of the loss function is very large, then the Fed should act as if it assigned weight on the inflation that is ten times more than suggested by the society's true loss function. Under this biased objective function, the optimal control policy reacts to inflation, performs better when investors are learning in the model, anchoring inflation expectations.

In a nutshell, low weights on unemployment and interest rate deviations from their target stabilize inflation and inflation expectations when investors learn. The OC policy reacts to return inflation back to its target rapidly after the shock. Hence, the public expectations anchor near the target, muting the impact of learning in the economy. In juxtaposition, if the weights on unemployment and interest rate deviations are high, then the OC policy very gradually returns the inflation rate to its target after a shock. These persistent deviations of inflation from its target can confuse the public as they have asymmetric information about the Fed's objective and the state of the economy.

Thus, I simulated another FRB/US model that shows the paths of the variables under OC policy with reduced weights of *unrate* and *ffr* relative to *pce_rate*. $w_u = w_r = 5$, and $w_\pi = 10$. Now, the federal funds rate lifts off six quarters later in 2017-Q4, than in the previous case, loosening the monetary policy. Consequently, *rgdpch* is also higher, and reaches the peak at 4 percent in 2015-Q3. *unrate* falls to 4.6 percent, also increasing *epop*. The easing policy also creates inflationary pressures, spiking both *pce_rate* and *corepce_rate,* but *eci_wage* doesn't significantly rise.

The ninth simulation measures the effects of a few aforementioned demand and supply shocks in the present from 2020-Q2 until 2023-Q4.[4] Invoking the Taylor rule in a demand shock, with *ffr* already in the ZLB today, the model projects *ffr* to lift-off in 2022-Q1 under $eciwag\_rate_{thresh} = 3.5\%$ and consensus baseline. *unrate* reaches ~4 percent under both the thresholds, which is approximately 1 percent below the baseline forecasts. Better labor market conditions heat the economy, raising the inflation rates to their target 2 percent levels by mid-2022. Yet, *eciwag_rate* lags behind and is only able to reach 2.8 percent by 2023-Q4.

In the tenth simulation, I forecasted the influence of a $20/barrel hike in the price of oil. In the wake of mounting inflationary pressures, these lift the *ffr* early in 2021-Q3 under both the thresholds. *unrate* initially soars at 13 percent in 2020, before it descends to 3.75 and 4.05 percent under $pce\_rate_{thresh} = 2.5\%$ and $eciwag\_rate_{thresh} = 3.5\%$, respectively in 2023-Q4. Unlike the sluggish pace of wage growth in the demand shock, *eciwag_rate* picks up faster, reashing 3.2 percent under the PCE inflation threshold at the end of the forecasting period. However, the growth is still tepid at 2.8 percent under the alternative threshold condition. We observe similar patterns in another supply shock − 2 percent decline in *lfpr*. However, the differential peaks of *unrate* observed in the three scenarios reveal inconsistencies. Under $eciwag\_rate_{thresh} = 3.5\%$, *unrate* reaches only 6.8 percent as opposed to 10 percent under $pce\_rate_{thresh} = 2.5\%$. Nonetheless, *epop* under both thresholds slump to the floor at ~0.55, when we would expect it to be lower under $pce\_rate_{thresh} = 2.5\%$. *rg10* climbs under $eciwag\_rate_{thresh} = 3.5\%$ and is ~1.6 percent higher than under the alternative threshold, reflecting lower demand for long-term yields. Thus, commercial banks increase long term rates on mortgages, auto-loans, credit cards etc, making investment harder, marginally diminishing *rgdpch* from 2021-Q3.

The twelfth simulation presents the impact of unanchored inflation expectations, drifting the inflation expectations upwards by 2 percent. Invoking the Taylor rule with the unemployment gap, *ffr* from both the thresholds climb at 2021-Q1 from the ZLB. As observed in previous projections, *ffr* rises faster under $eciwag\_rate_{thresh} = 3.5\%$, tightening the monetary policy. Thus, the inflation rates from the ECI wage rate threshold are lower than those in the PCE rate threshold. Additionally, *rgdpch* and *epop* under the former grows slower after the rate hike relative to *rgdpch* in the latter threshold. Next, I have modeled the paths, separately under the inertial Taylor rule and Taylor rule, keeping the thresholds in place. Most variables behave somewhat similar to the trajectories observed above, save for *ffr*, as it lifts a quarter later in 2021-Q1.

---

[4] I inserted the consensus baseline forecasts from the Summary of Economic Projections released by the FOMC on June 10, 2020.

Furthermore, I modeled the impulse response functions to reflect the paths of changes in the variables if the Fed adopts a more aggressive stance of the Taylor rule monetary policy by changing its parameters. Roberts and Reifschneider (2006) explored a few proposals to reduce the impact of ZLB on interest rates, and how to affect the investors' expectations of future monetary policy when *ffr* is at ZLB. The Fed is concerned about the ZLB because weak aggregate demand warrants the Fed to lower the short-term interest rates. Coupled with the expectation that the *ffr* will remain at low levels until the economy restores to pre-crisis levels, the long-term interest rates also decline, stimulating the aggregate demand. However, the Fed is constrained to provide further stimulus via this channel once the *ffr* hits ZLB. Nonetheless, the Fed can sway the expectations of future short-term interest rates, for at least the time far into the future until the policy will not be constrained by ZLB. Forward guidance through FOMC's statements about the future plans for setting the *ffr* can influence the investors' expectations as they observe the historical workings of the Fed's reactions. Also, the expectations hypothesis of the term structure of interest rates states that the Fed's influence will alter the real rates on bonds, if the investors deem its statements to be credible, uplifting the sagging economy. This strategy is effective in the FRB/US model as the real long-term interest rates in the present, functioning via the cost of capital, directly affect the current spending, and paths of long-term rates. Moreover, they indirectly affect inflation and real activity in the model as they influence the asset prices and real exchange rates.

Krugman (1998) recommended either to permanently raise the inflation target or do so for a long time given that in recent times, inflation rates have predominantly been below the Fed's target. In spite of the easing policy, the last time *pce_rate* breached the 2 percent target was in March 2018. Alternatively, Reifschneider and Williams (2000) proposed adopting a more aggressive monetary policy to subdue the detrimental effects of ZLB. I attempted to model this aggressive approach by replacing the current coefficients of $(y_t - y_t^P)$ − output gap and $(\pi_t - \pi^*)$ − inflation gap, with larger coefficients in the Taylor rule.

The original Taylor rule is: $R_t^T = r_t^{LR} + \pi_t + 0.5\,(\pi_t - \pi^*) + 0.5\,(y_t - y_t^P)$

I modified it to $R_t^T = max\{\, r_t^{LR} + \pi_t + 2\,(\pi_t - \pi^*) + 2\,(y_t - y_t^P),\ 0\}$ and simulated the effects of a strong negative demand shock. Raising the coefficients each from 0.5 to 2 indicate aggressive response to inflation and output gaps. The *max* function restraints the nominal interest rates $R_t^T$ to the ZLB. The hypothesis is that this aggressive strategy will lessen the public's expectations about the path of future *ffr* and other short-term treasury bills, albeit they are currently nearly 0. Promising an easy policy in the future will raise inflation expectations.

The graphs of the impulse response functions in the fourteenth simulation show the paths of the macroeconomic variables under the original and modified versions of the Taylor when the Fed adopts $pce\_rate_{thresh} = 2.5\%$ under ZLB. Under the modified version, *ffr* lifts in 2024-Q2, which is two years later compared to the paths prescribed by the original rule. From 2023-Q3, *rgdpch* surpasses the GDP growth rate observed under the consensus baseline and the benchmark policy − unconstrained standard Taylor rule. The aggressive policy more efficaciously checks the inflation decline, which slumps the *rg10* by a greater extent, stimulating the economy. We can attribute the smaller inflation effect to the investors' beliefs that the Fed when unconstrained, will aggressively seek to anchor the inflation rate to its target rate. As the expected ex-ante inflation rate changes the current *pce_rate,* this accommodative mechanism boosts *pce_rate* albeit *ffr* is bounded at 0. Alternatively, *eciwag_rate* crosses 3.5 percent in 2026-Q2, evident from the slow and steady adjustments of wages to faster changes in price.

Finally, I simulated the effect of a temporary boost in government spending in the advent of a negative aggregate demand shock provided that the Fed adopts an accommodative monetary transmission channel for prolonged periods. For that, I have assumed that the Fed adheres to the policy paths prescribed the modified Taylor rule described above. Unlike in the previous simulation when *ffr* lifted in 2024-Q2, now it lifts early in 2023-Q3. As opposed to the decline of *unrate* without the additional government expenditure, the stimulus lowers *unrate* by approximately 0.4 percent in the modified version, marginally raising *epop* as well. *eciwag_rate, pce_rate* and *corepce_rate* slightly rise as the discretionary fiscal stimulus shock raises the fiscal multiplier.

The upside of an aggressive Taylor rule is that the Fed could enforce it even in normal circumstances when the ZLB does not constrain the typical open market operations. Thereby, policymakers would gain credibility and establish a reputation of aggressively responding to movements in inflation. The downside is that very likely, agents beyond those in the financial markets (such as households and firms), gradually learn the impact of monetary policy on inflation as they observe the ex-post values, rather than the ex-ante inflation. Hence, if some agents don't alter their expectations instantaneously after the Fed announces a new policy, but instead rely on the average historical trajectory of the benchmark rates, then the benefits of the announced policies will be substantially curtailed. Furthermore, imperfect credibility by the investors during ZLB can be problematic as the inflation may overshoot, building up financial risks, and un-anchoring inflation expectations (Brainard 2017). As a corollary, before resorting to new policy regimes, policymakers should be thoroughly confident that the proposed changes would work, albeit investors are only partially credible about its functionalities and effectiveness.

## Why does $eciwag\_rate_{thresh} = 3.5\%$ Tighten the Monetary Policy?

After imposing two supply shocks, a negative aggregate demand shock, unanchored inflation expectations and unexpected hike in the federal funds rate, the findings conclude that the ECI wage growth rate of 3.5 percent is still not accommodative enough, and wage growth higher than 3.5 percent may be necessary. This is likely because of the way the equations in the FRB/US model have been formulated with respect to the PCE inflation rate and the variables derived from it, such as the expectations of PCE inflation rate in the next quarter. For instance, the monetary policy reaction functions and the loss function in OC policy consist of an inflation gap based on the PCE inflation rate. One might consider the possibility of incorporating a wage growth gap in the reaction functions, but that would require us to know the target wage growth rate. Besides, we would have to check the feasibility of incorporating *eciwag_rate* in those reaction functions as *eciwag_rate* is not a proxy variable for *pce_rate* and cannot be substituted for *pce_rate* without gauging the true relationship between the dependent variable in question and *eciwag_rate,* and the biases and errors it produces.

Another potential reason could be the inherent property of price and wage rigidity driving *eciwag_rate.* A key mechanism that monetary policy affects the economy is price rigidity. A form of Phillips curve, known as the New Keynesian Phillips Curve (NKPC) explains the linkage between inflation, price rigidity and movements in the real economy as it associates inflation to factors such as production and utilization costs. However, a diverse array of papers have demonstrated that NKPC fails to capture the persistence of inflation, overstating the function of expectations in setting prices; thus prices are excessively rigid. A model that cannot satisfactorily account for inflation's persistence is of questionable value for forecasting. Furthermore, amplifying the magnitude of price rigidity may inadvertently overstate the Fed's relevance in determining real outcomes, possibly distorting the role of monetary policy in macro-stabilization.

### Discussion: Should the FOMC Adopt a Wage Growth Rate Threshold Higher than 3.5 Percent?

The results of the simulations indicate that establishing an ECI wage growth threshold at 3.5 percent generally tightens the monetary policy as the federal funds rate lifts-off earlier than if the PCE inflation rate threshold of 2.5 percent is in place. Hence, should the Fed consider raising the wage growth rate threshold to values greater than 3.5 percent to check if this aggressive stance boosts the sagging economy? In retrospect, the episodes of policy rate hikes from 2015 and the trajectory of labor utilization and price variables suggest that raising the wage growth rate threshold is feasible without instigating fears of overshooting the inflation rate.

In December 2015, the Fed lifted the interest rate by 25 bps from the ZLB for the first time since 2008-Q4, albeit *unrate* was still elevated at 5 percent, and *pce_rate* was barely crawling at 0.35 percent. Hiking the policy rates even before the inflation rate picked up steam stands as testimony to the fact that the Fed was unwilling to aggressively experiment with letting *unrate* fall even further and give more room for inflationary pressures. Failing to aggressively target a lower *unrate* is a major reason for the anemic growth in wages. Whilst the Fed raised it by another 25 bps after one year in December 2016, it lifted the rates seven times, each by a quarter-point within two years in 2017-18.

Ideally, the Fed should hike the rates only to inhibit the economy from overheating i.e if there are fears that rapid growth and declining unemployment will ignite fast wage growth and overshoot inflation. Nevertheless, no such signs were prevalent for the Fed to diagnose the economy with such premature rate hikes. Perhaps, the concerns of the wage-price spiral were high - tighter labor market conditions would increase the bargaining power of laborers as they would be able to demand higher wages and enhanced fringe benefits, spiking inflation as firms would pass on the higher labor costs to the consumers. So, traditionally, the higher labor costs arising due to higher wages are a precursor to inflationary pressures, that the Fed should counteract.

Theoretically, a wage-price spiral causes workers in full employment to target growth in real wages (wages adjusted for inflation) which equates to productivity growth. Moreover, they can bargain to accommodate their nominal wage demands by acquiring wage raises at the rate of inflation in the previous year. When *unrate* is greater than NAIRU, then higher unemployment hamstrings them to earn their true aspired real wages; thus workers target real wage growth that falls short of productivity. Alternatively, if *unrate* sinks below NAIRU, then workers shoot for higher real wages that outgrow productivity. Whilst unattainable, as productivity is the ceiling on the maximum yield an economy can generate, their nominal wage demand sets in the wage-price spiral in motion − workers earn nominal wage hikes, but employers pass over the higher labor costs to consumers. This offsets (and erodes) the real wage growth, continuing the feedback loop. To break the wage-price spiral, policymakers slow the economy's pace by indirectly attenuating the bargaining powers of workers as *unrate* rises.

Yet, higher *eciwag_rate* did not materialize although *unrate* dipped below NAIRU as shown from the graph above. We don't know the *unrate* low enough to jump *eciwag_rate* and *pce_rate*. Nor do we know the *unrate* level for which the *pce_rate* will not just soar but will gain momentum and accelerate if *unrate* falls below NAIRU. *unrate* breached NAIRU for the first time in January 2017 since December 2007, but paradoxically *pce_rate* diminished from 2.01 to 1.53 in June 2017, and *eciwag_rate* barely budged from 2.38 in January 2017 to 2.61 percent seven months later. Whilst the economy plunged into a cataclysmic wage-price price spiral in the late 1960s and 1970s as prices for consumers surged from 1.6 to 13.5 percent, the wage-price spiral is dormant in the 21st century. To corroborate this, the OLS regression on the quarterly data on *eciwag_rate* and *pce_rate* from March 2002 to June 2020 demonstrated that a 1 percent increase in *eciwag_rate* augments the *pce_rate* by merely 0.461 percent, and the p-value was statistically significant.

$$pce\_rate = 0.6488 + 0.461 \, eciwag\_rate$$

$$(0.463) \quad\quad (0.182)$$

Given the tenuous connection between wage growth and PCE inflation rate in recent years, it is far-fetched that a wage-price spiral would usher in, making the tightening policy redundant. There are miniscule dangers of overshooting inflation in the present day. On the contrary, overshooting inflation should, in fact, be a part of the Fed's normalization policy, and not a problem to avoid.

## Conclusion

I have examined the Evan's Rule in the context of state-based forward guidance that the FOMC adopted in the aftermath of the Global Financial Crisis. After reviewing the drawbacks in the measures of unemployment rate and inflation rate, I proposed an alternative threshold of the ECI wage rate, and calibrated the $eciwag\_rate_{thresh}$ equivalent to $pce\_rate_{thresh} = 2.5\%$.

Transforming the standard Taylor rule to an unconstrained version loosens the monetary policy as it prescribes a later liftoff of the federal funds rate from the ZLB. Further research can explore the combination of "lower for longer" (L4L) $eciwag\_rate_{thresh} > 3.5\%$. L4L was an approach proposed by Bernanke et. al (2019) wherein the Fed would keep the rates "lower for longer" when the ZLB is hit. A completely credible L4L policy can extensively ameliorate the constraints imposed by ZLB. Additionally, it diminishes bond yields and increases inflation expectations, which reduce real long-term interest rates. Thus, markets are optimistic of ex-ante growth, encouraging consumers and producers to spend more during ZLB episodes.

# Appendix

The following pages contain the results of the FRB/US Simulations.

1. Macroeconomic Effects of a Negative Aggregate Demand Shcok
(VAR Expectations; Policy = rfftay)
(ZLB and Thresholds Imposed)

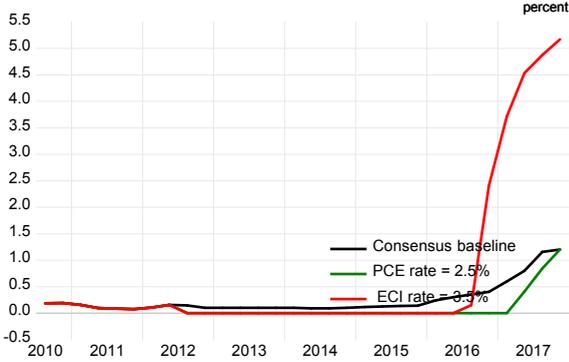
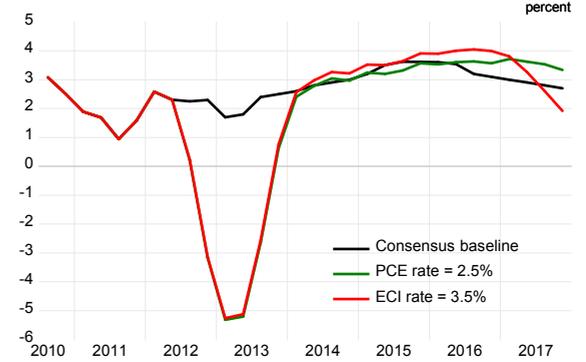
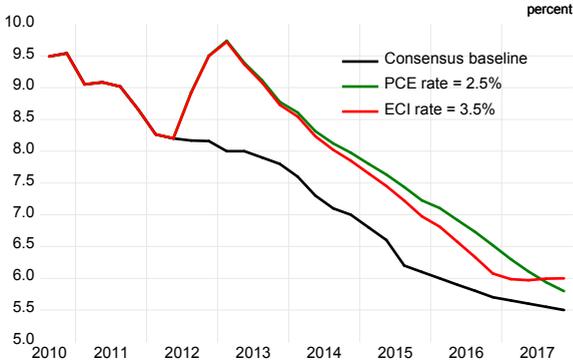
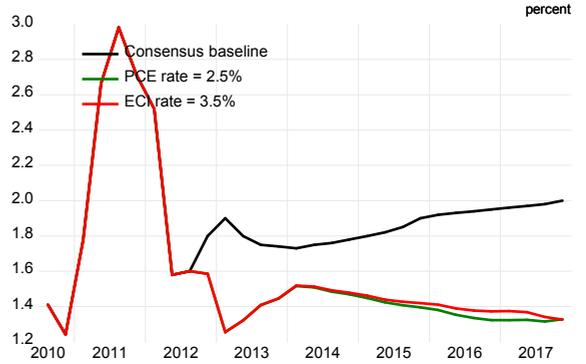
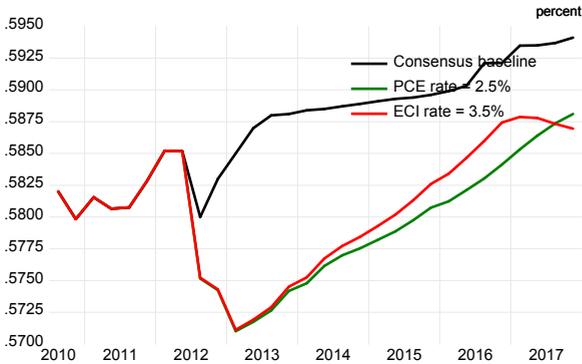
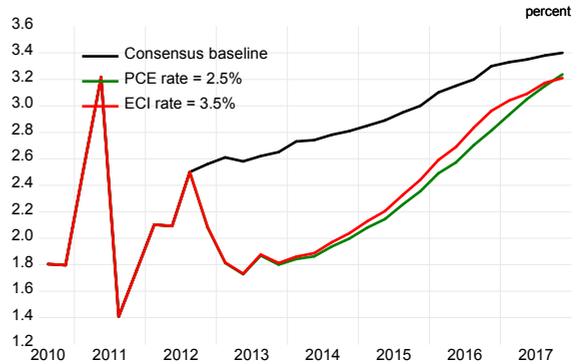
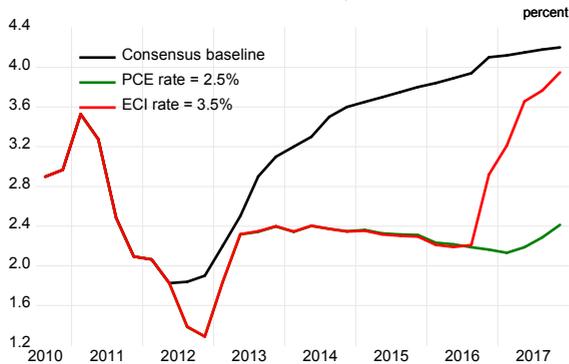
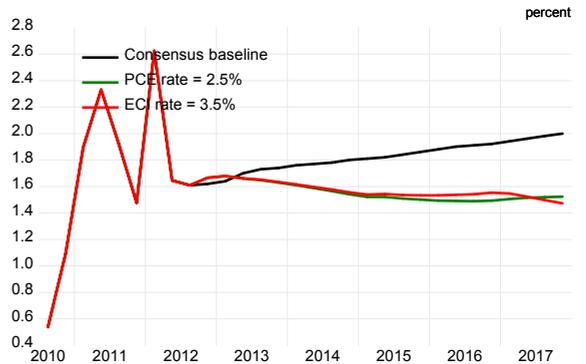

## 2. Macroeconomic Effects of $20/Barrel Higher Oil Prices
### (VAR Expectations; Policy = rfftay)
### (ZLB and Thresholds Imposed)

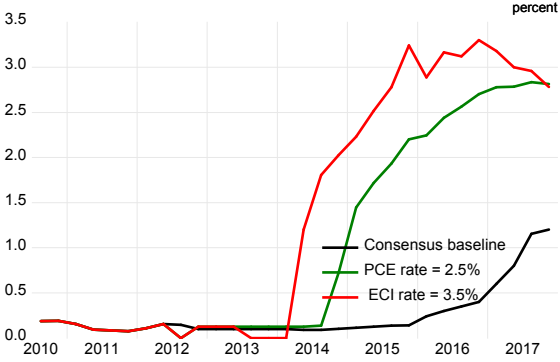

Federal Funds Rate

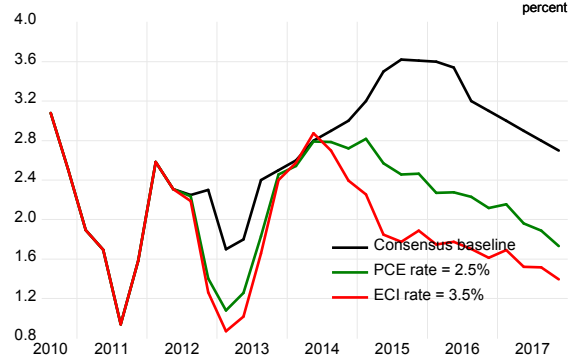

Year to Year % Change in Real GDP

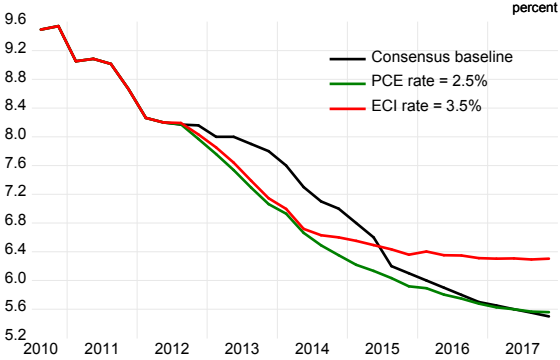

Unemployment Rate

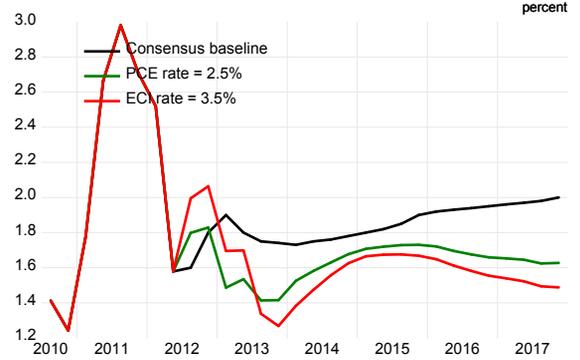

PCE Inflation Rate (4-Quarter)

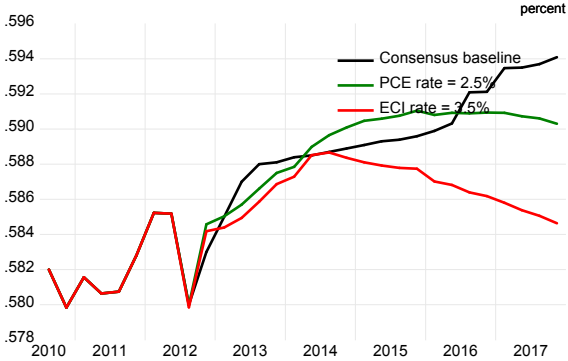

Employment to Population Ratio

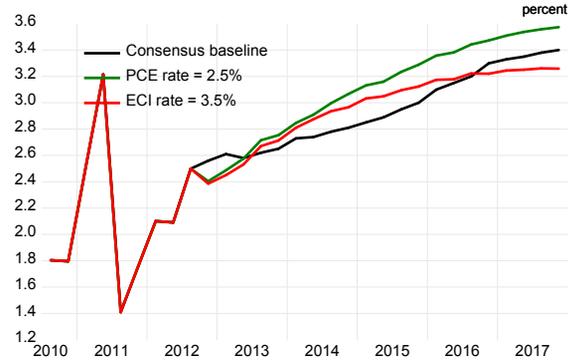

Annualized rate of growth of EI hourly compensation

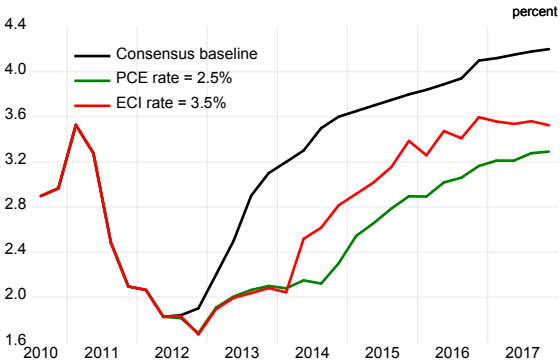

10-Year Treasury Rate

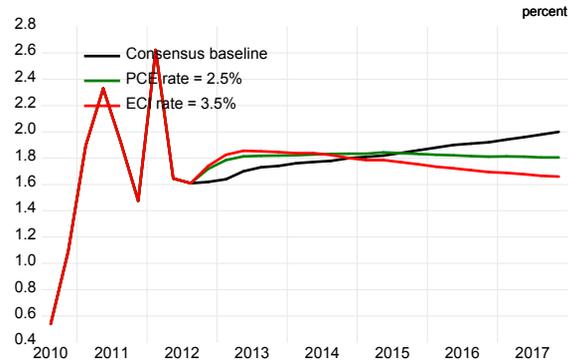

Core PCE Inflation Rate

3. Macroeconomic Effects of a Lower Labor Force Participation Rate
(VAR Expectations; Policy = rfftay)
(ZLB and Thresholds Imposed)

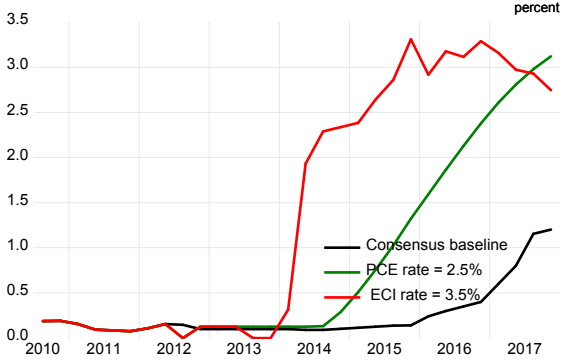
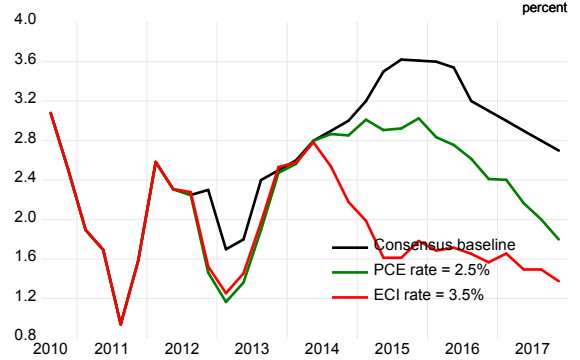
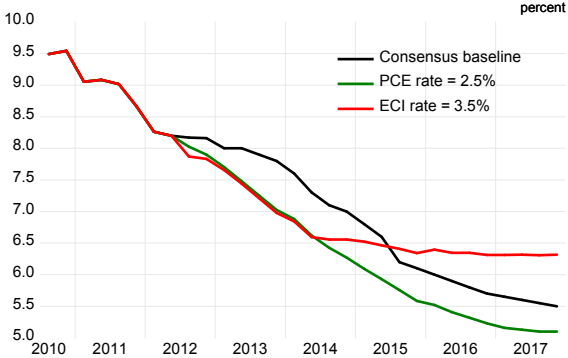
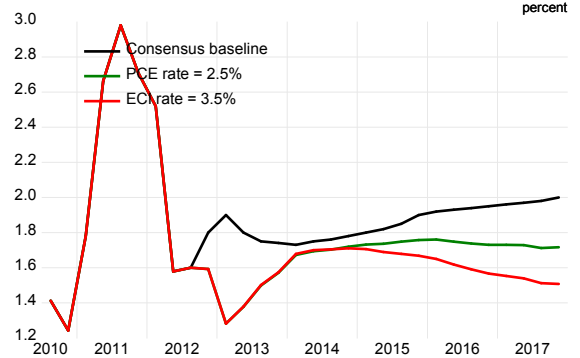
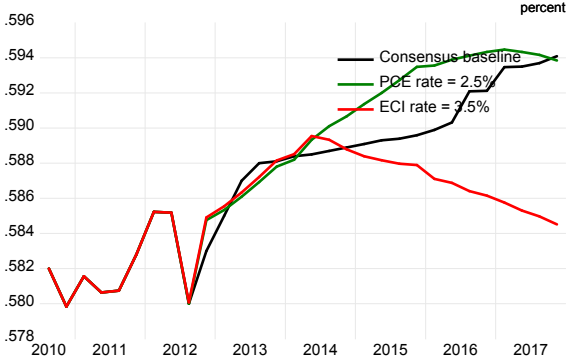
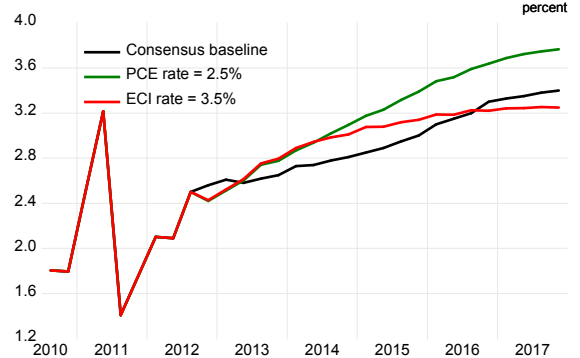
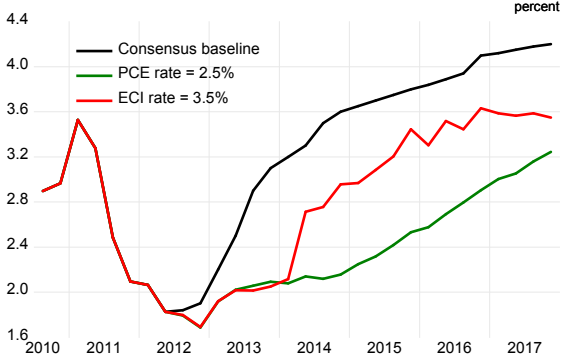
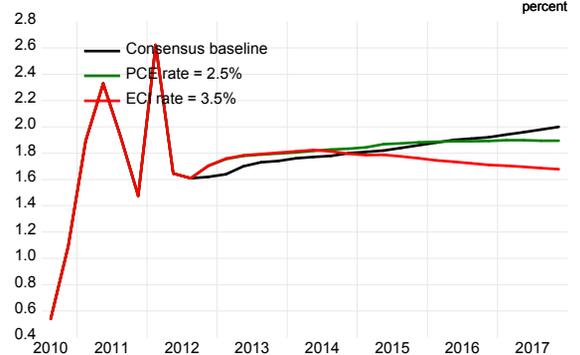

4. Macroeconomic Effects of Unanchored Inflation Expectations
(VAR Expectations; Policy = rfftay)
(ZLB and Thresholds Imposed)

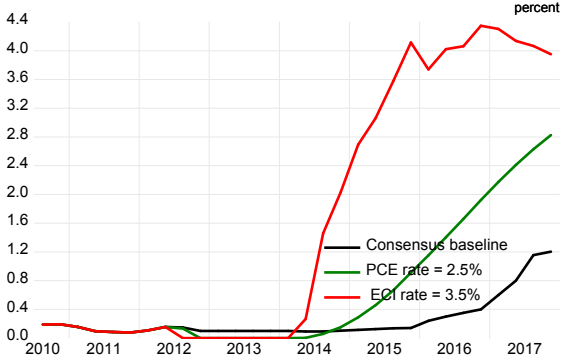
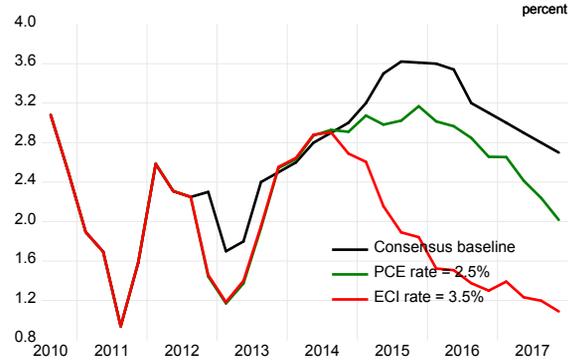
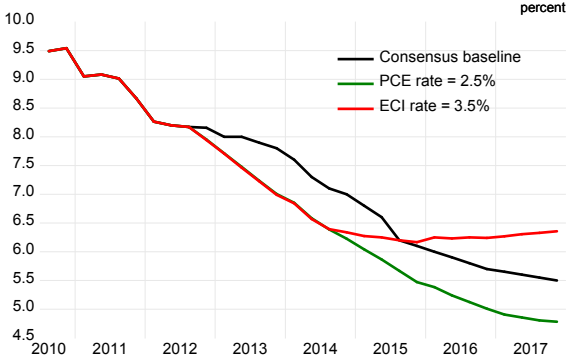
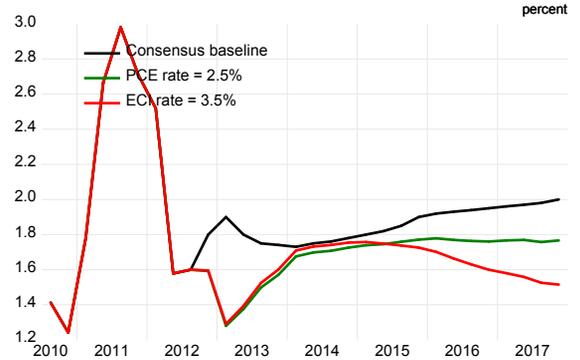
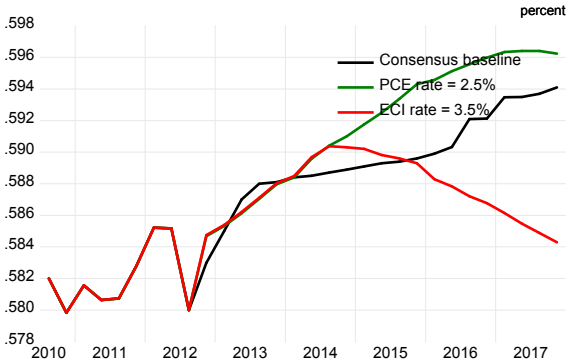
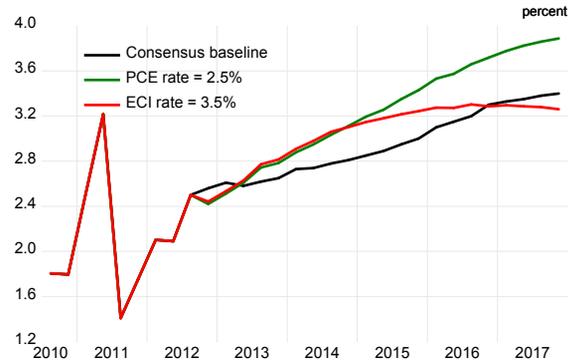
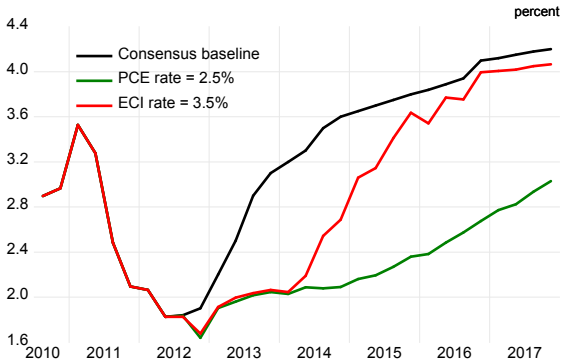
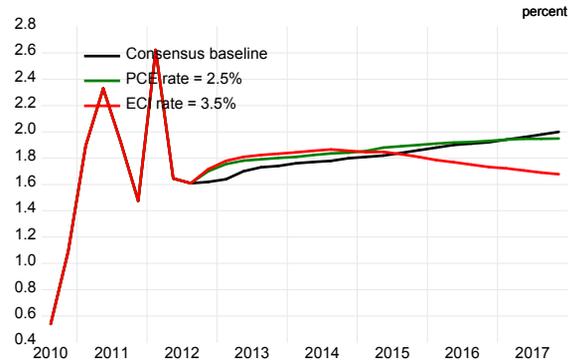

5. Macroeconomic Effects of the Taylor Rule with the Unemployment Gap
(VAR Expectations; Policy = rfftlr)
(ZLB and Thresholds Imposed)

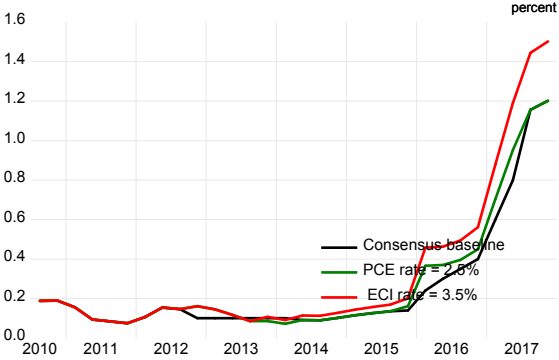
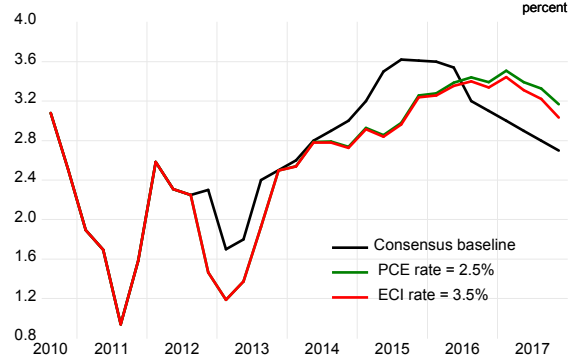
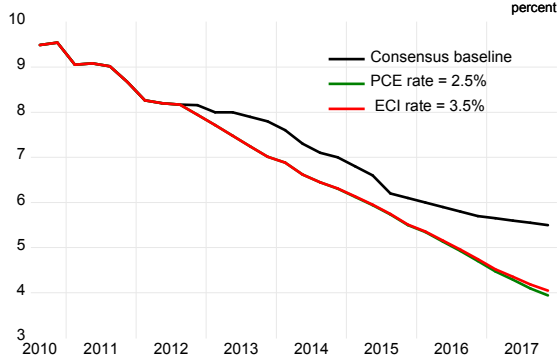
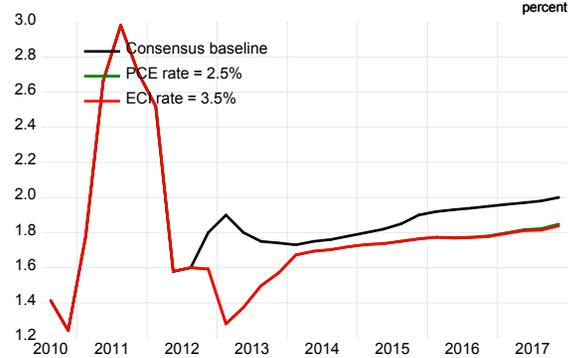
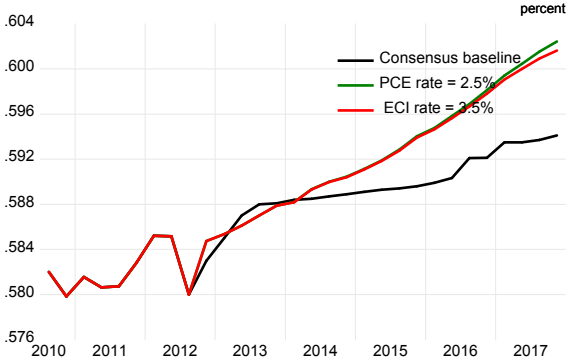
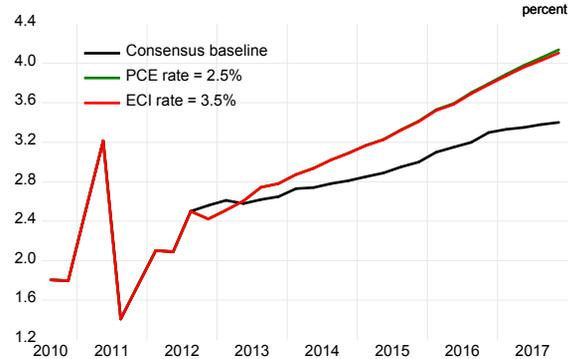
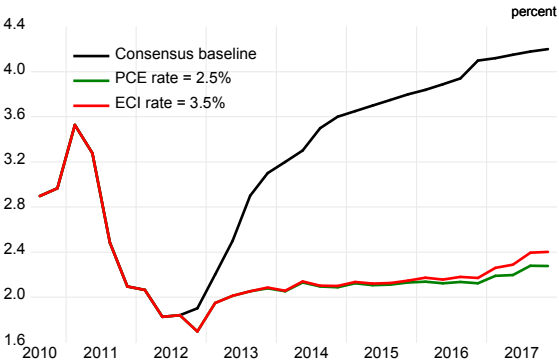
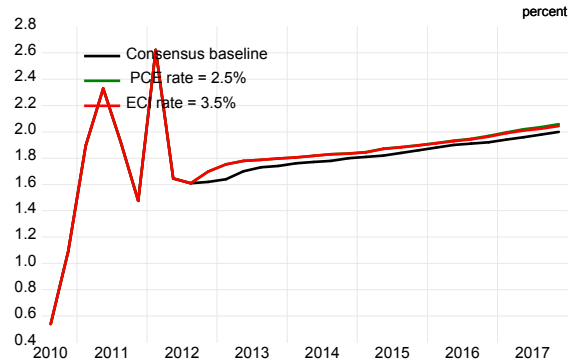

6. Macroeconomic Effects of an Upward Federal Funds Rate Shock

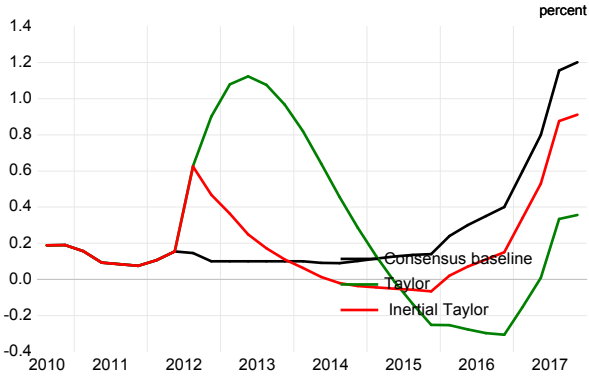

Federal Funds Rate

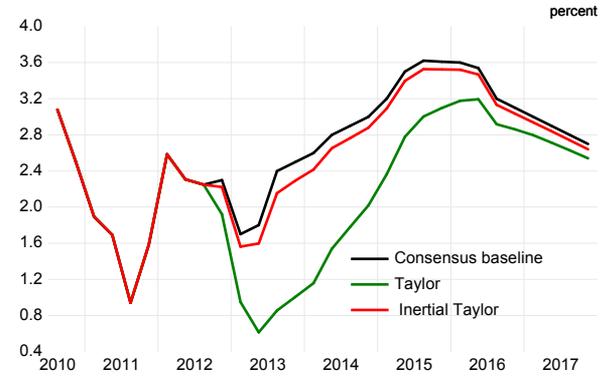

Year to Year % Change in Real GDP

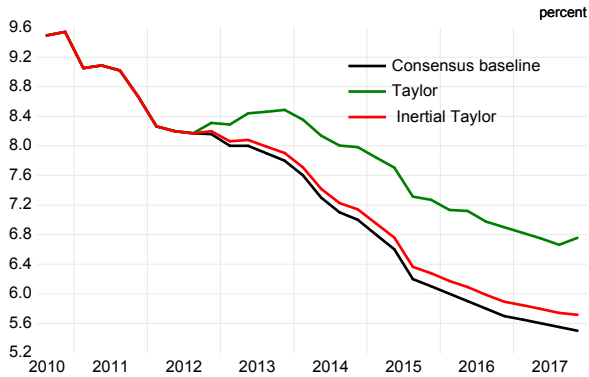

Unemployment Rate

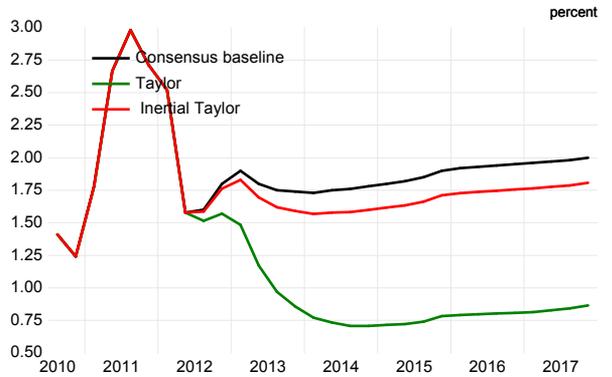

PCE Inflation Rate (4-Quarter)

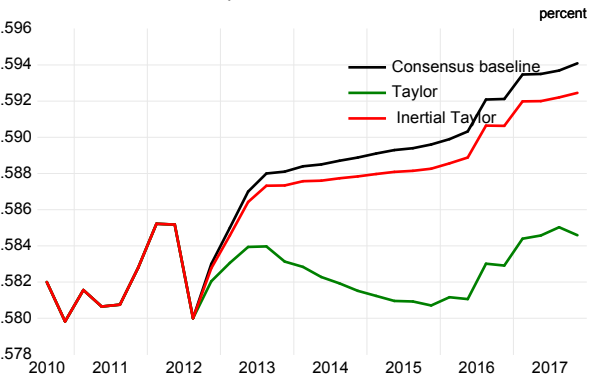

Employment to Population Ratio

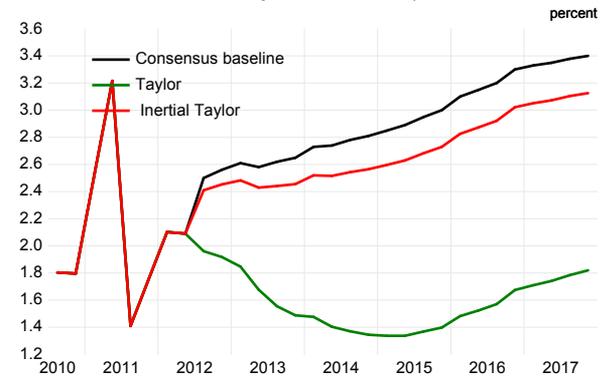

Annualized rate of growth of EI hourly compensation

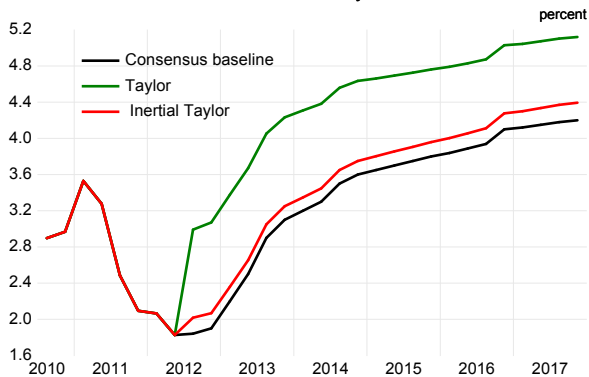

10-Year Treasury Rate

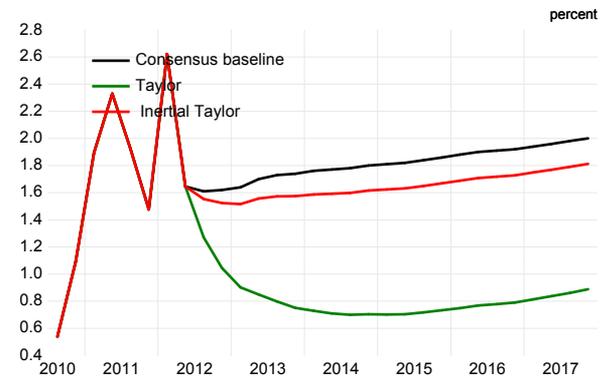

Core PCE Inflation Rate

## 7. ZLB Imposed

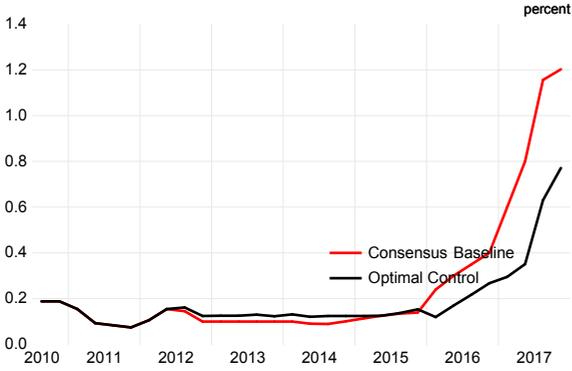
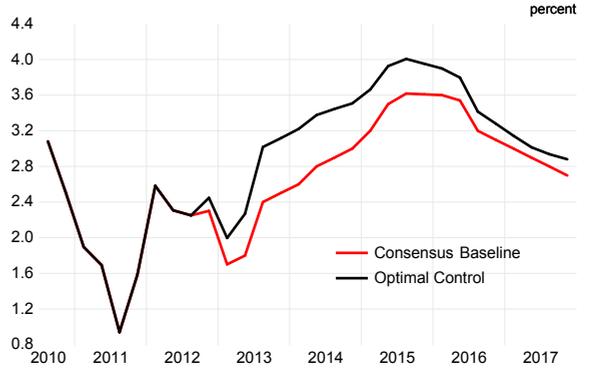
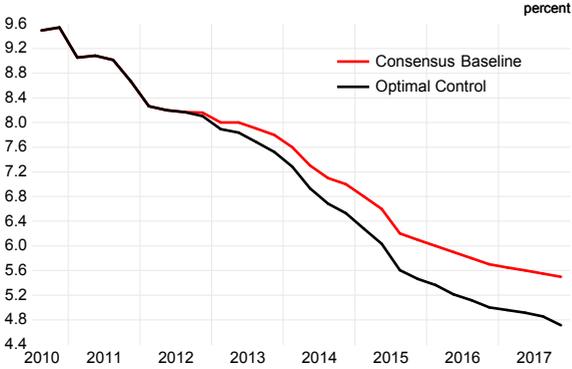
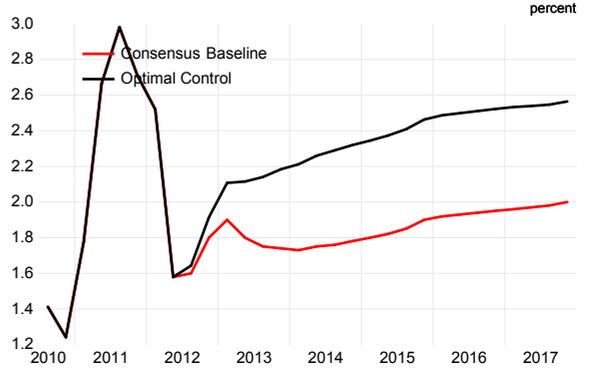
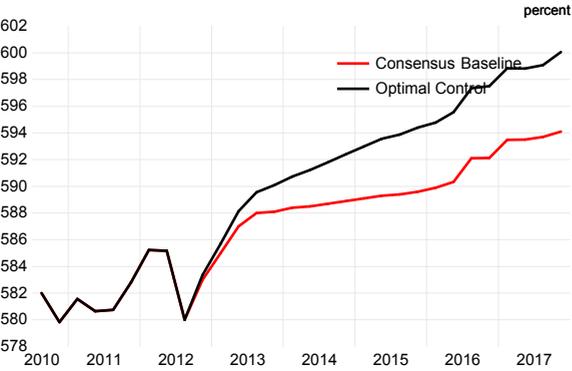
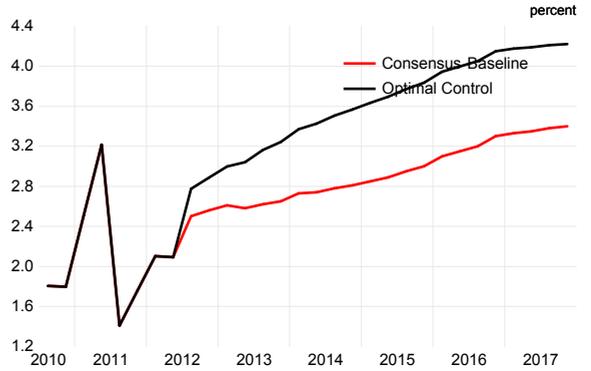
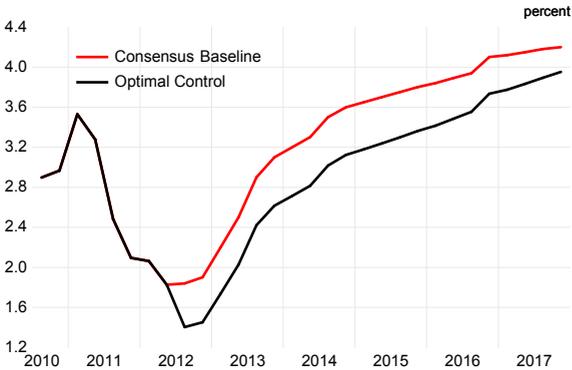
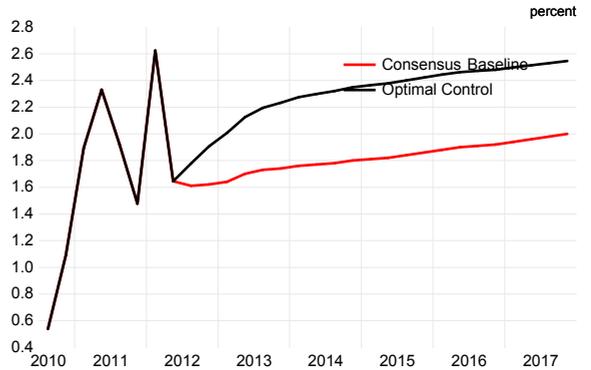

8. ZLB Imposed

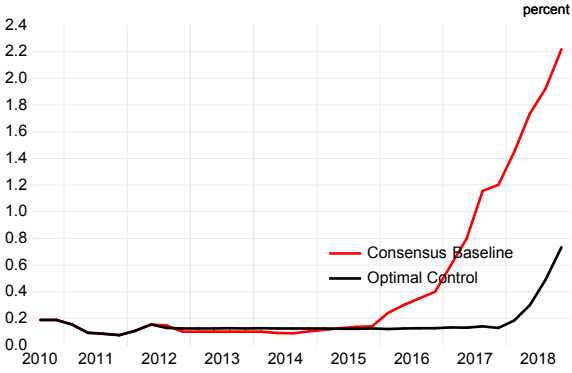
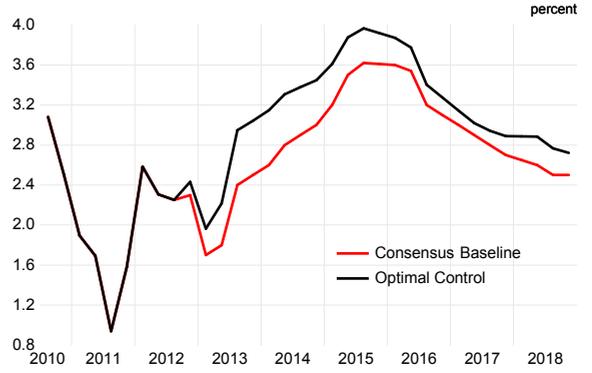
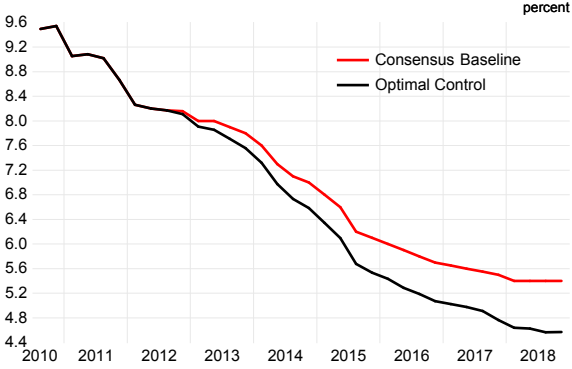
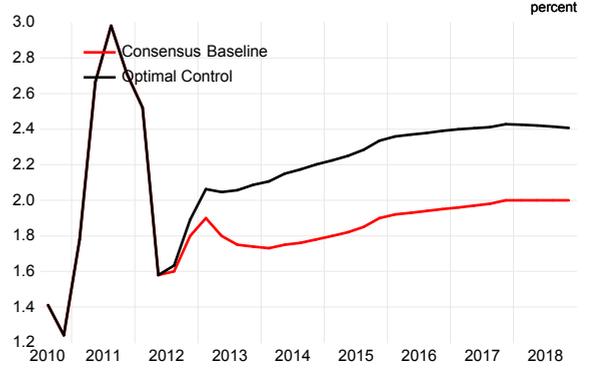
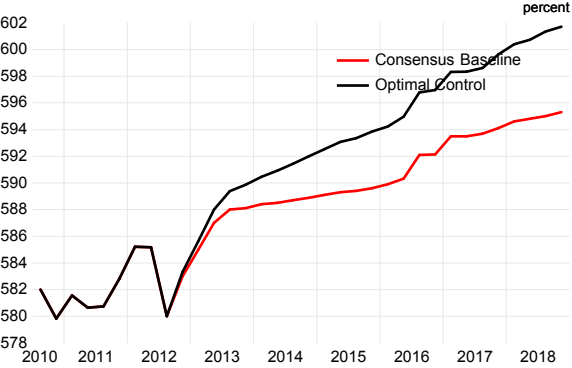
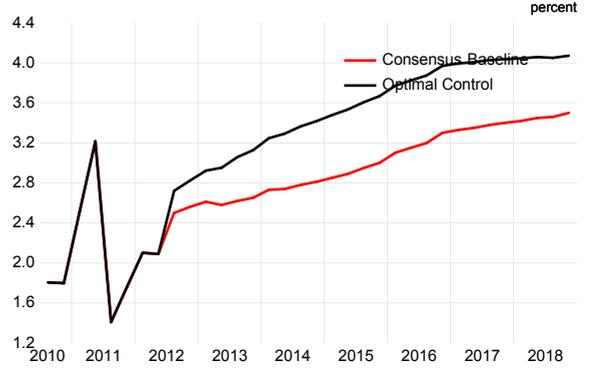
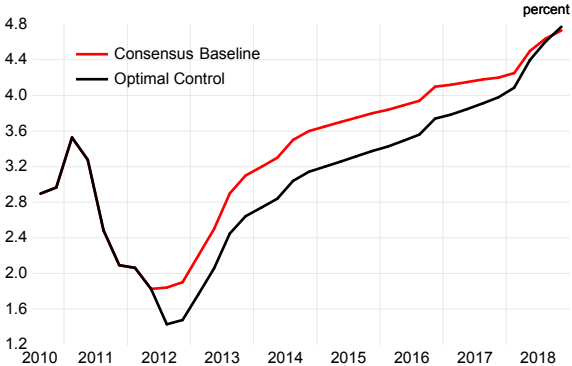
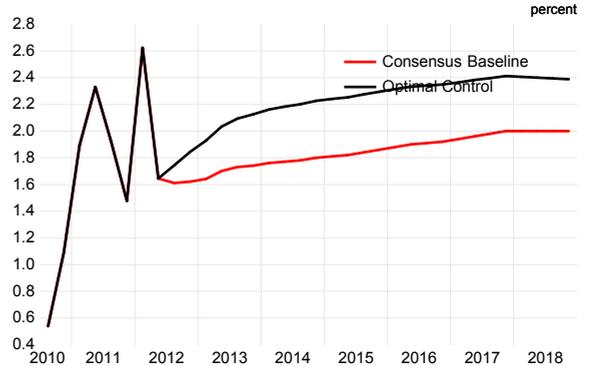

9. Macroeconomic Effects of a Negative Aggregate Demand Shock
(VAR Expectations; Policy = rfftay)
(ZLB and Thresholds Imposed)

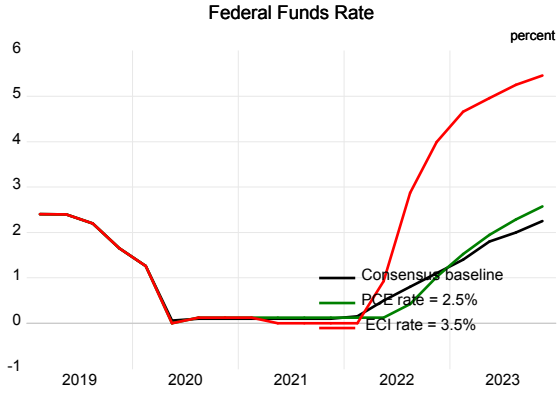
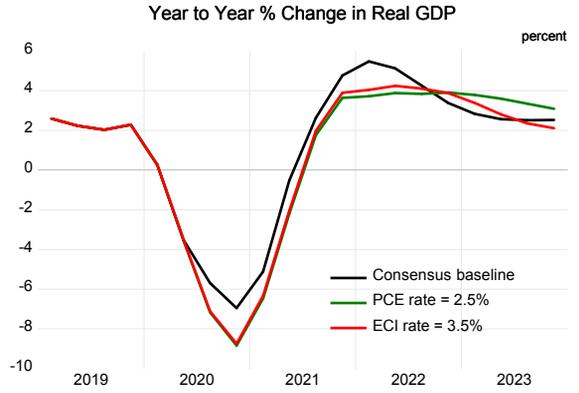
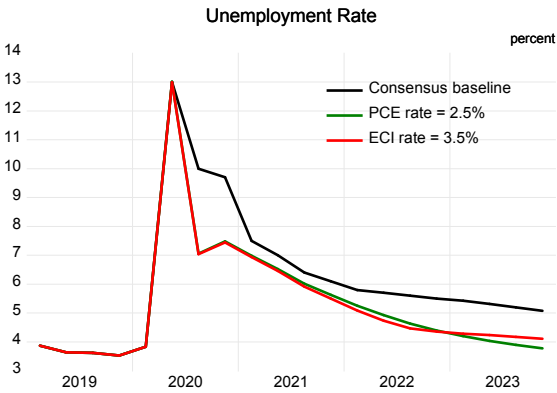
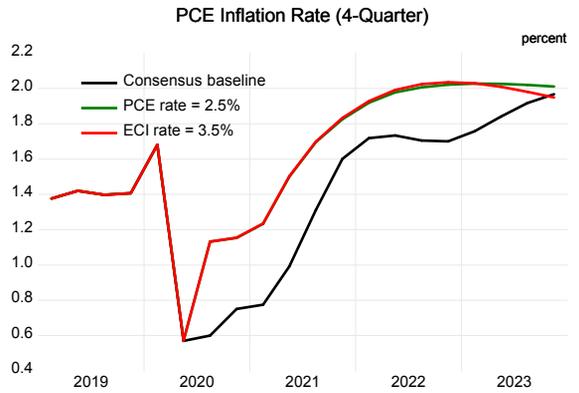
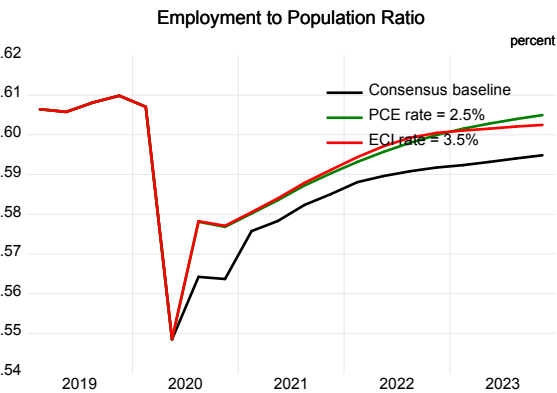
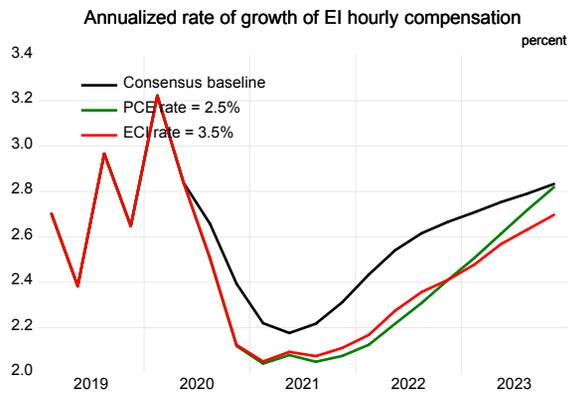
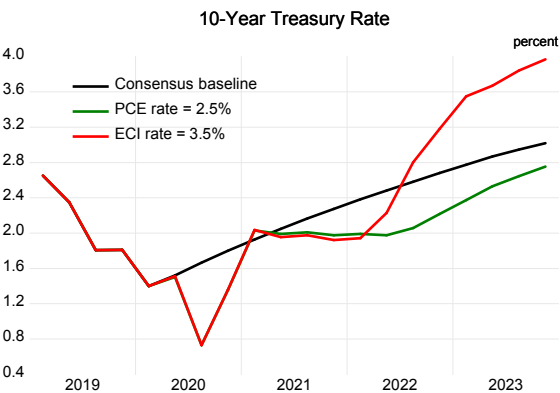
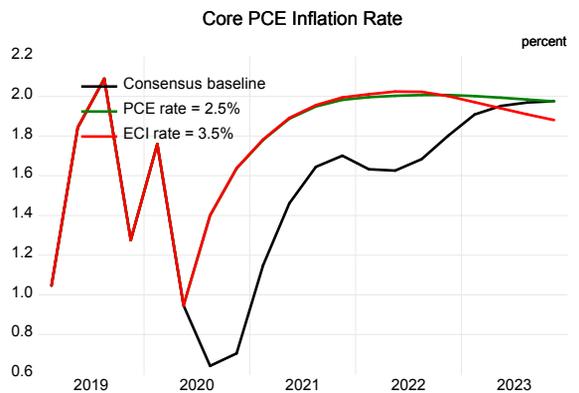

10. Macroeconomic Effects of $20/Barrel Higher Oil Prices
(VAR Expectations; Policy = rfftay)
(ZLB and Thresholds Imposed)

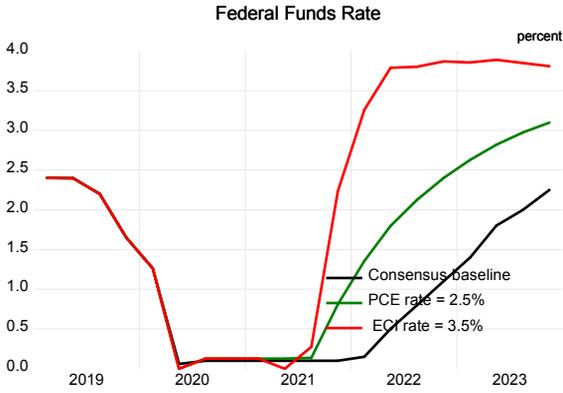
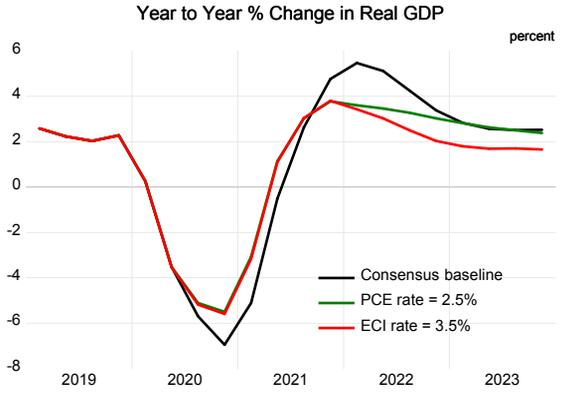
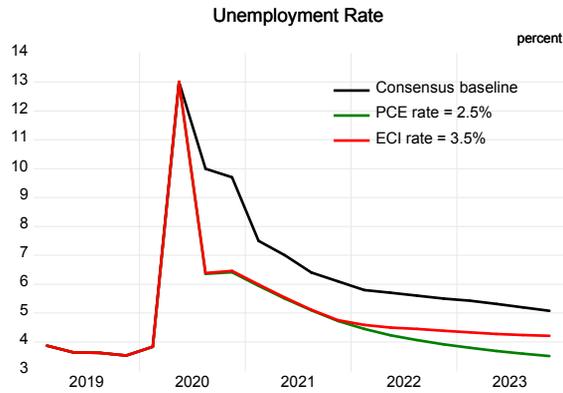
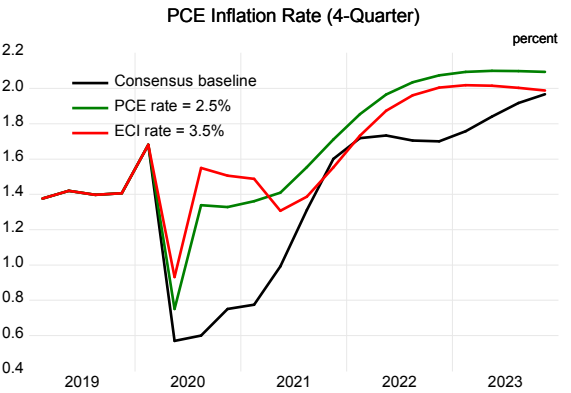
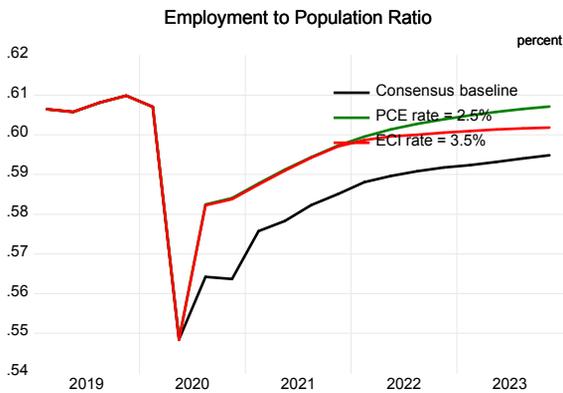
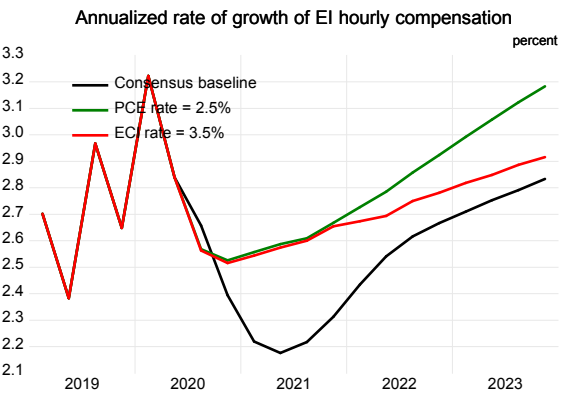
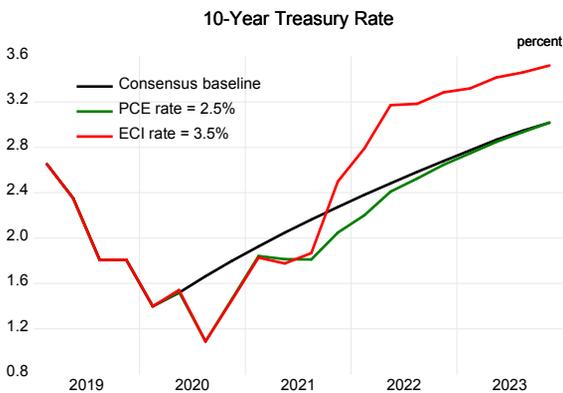
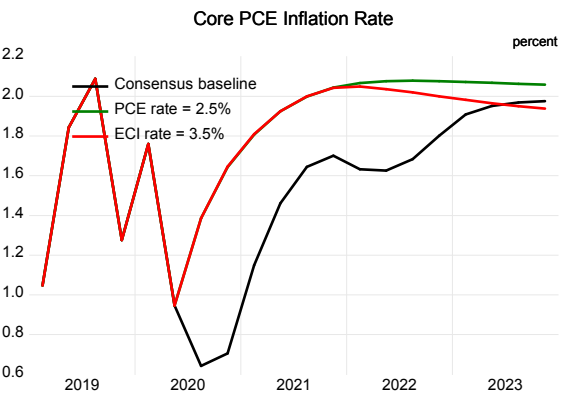

11. Macroeconomic Effects of a Lower Labor Force Participation Rate
(VAR Expectations; Policy = rfftay)
(ZLB and Thresholds Imposed)

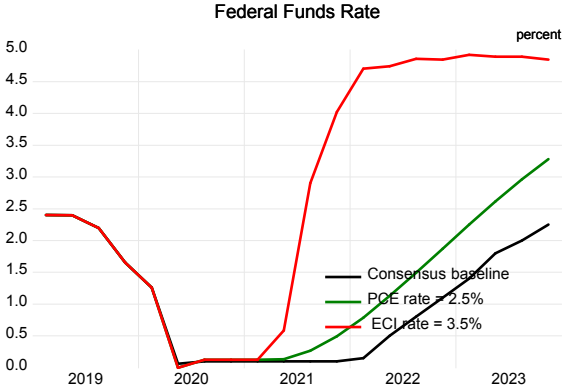
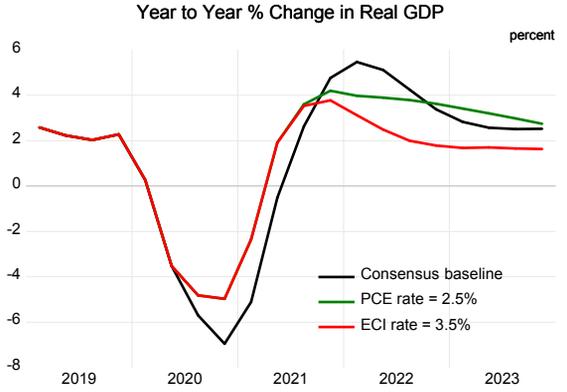
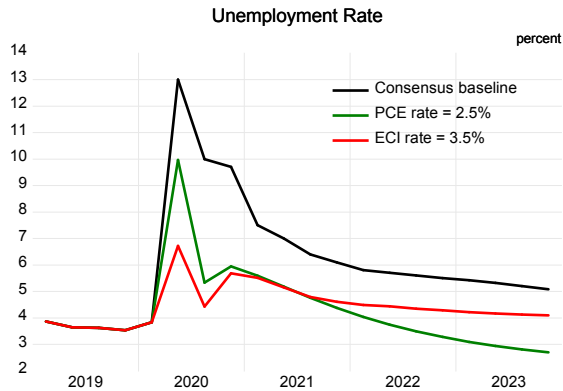
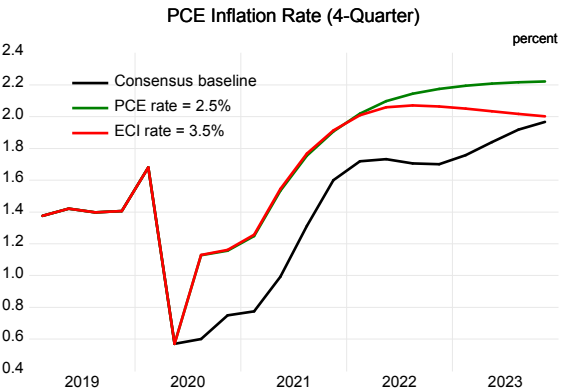
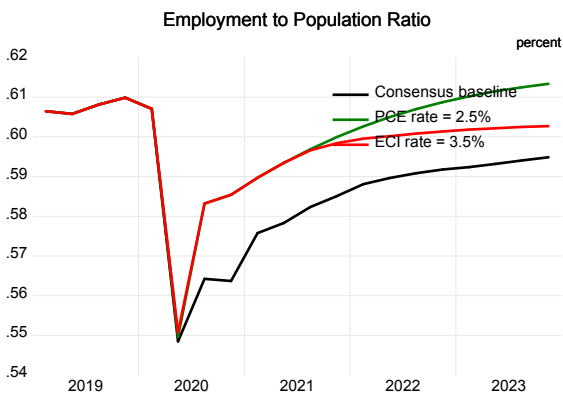
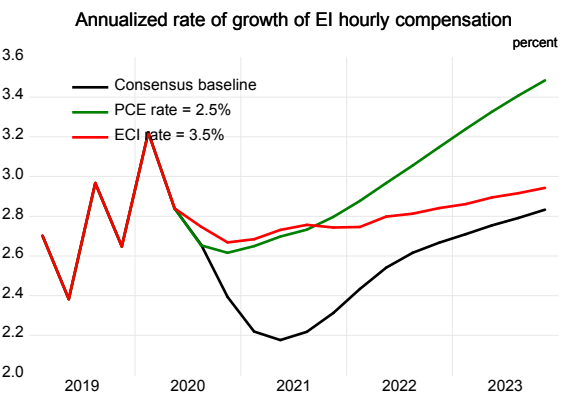
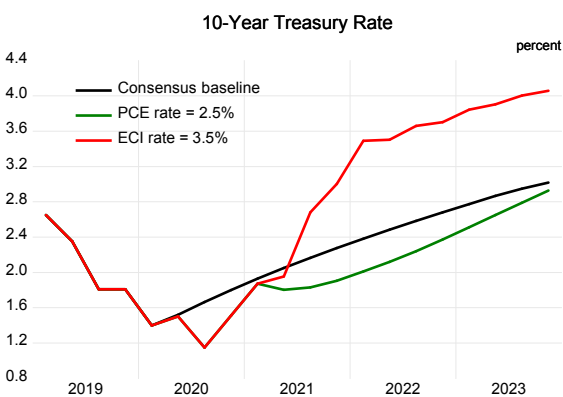
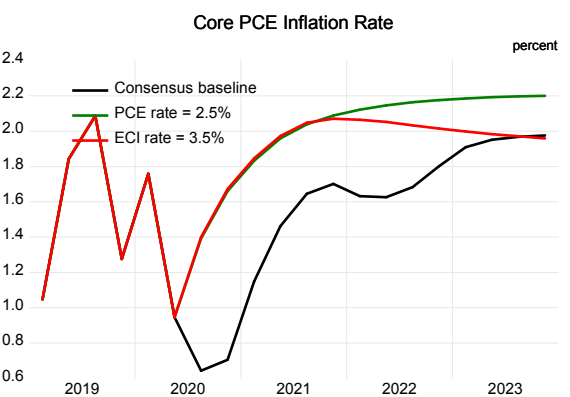

12. Macroeconomic Effects of Unanchored Inflation Expectations
(VAR Expectations; Policy = rfftlr)
(ZLB and Thresholds Imposed)

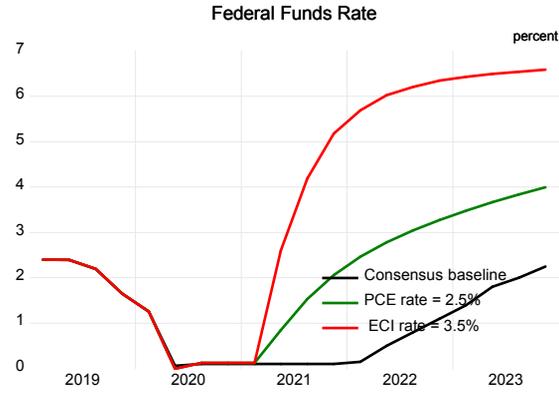
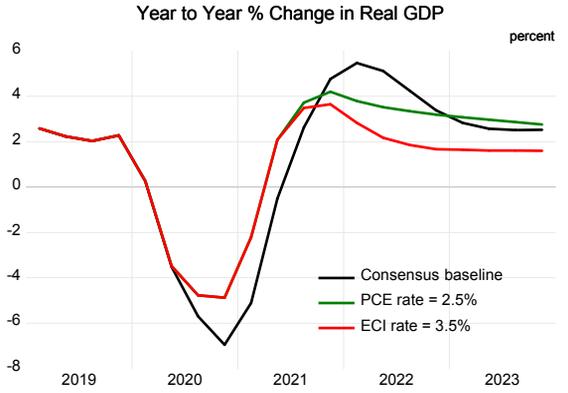
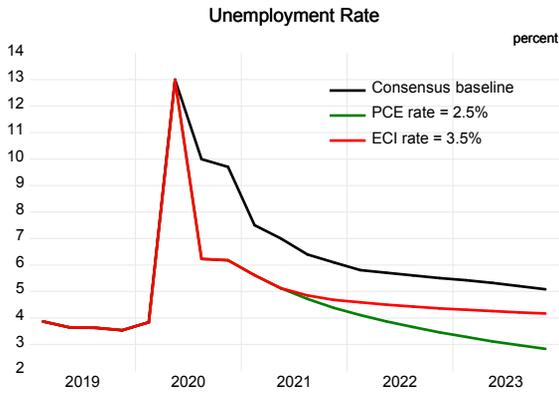
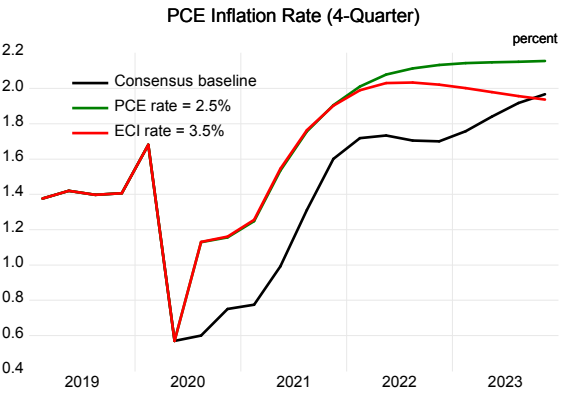
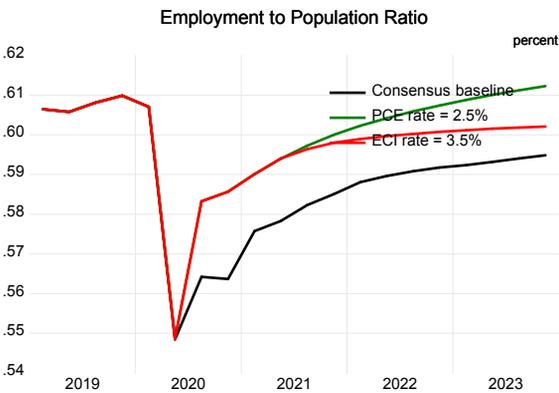
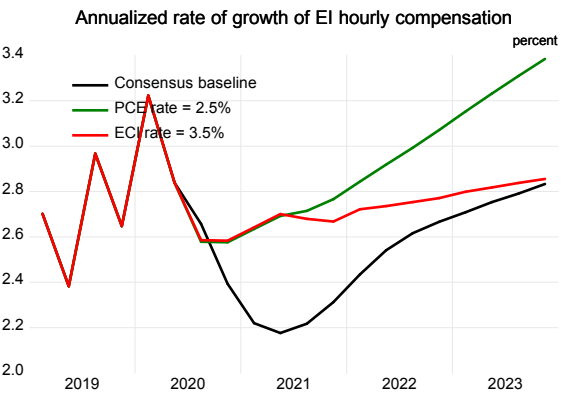
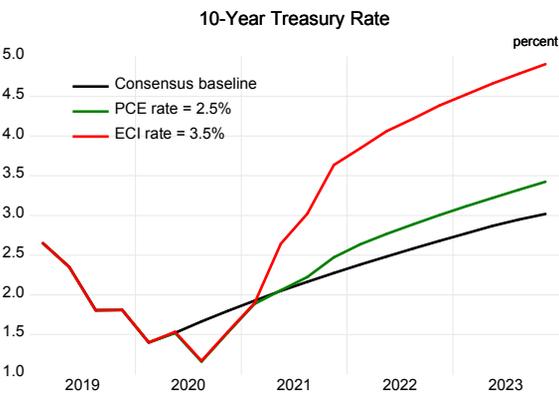
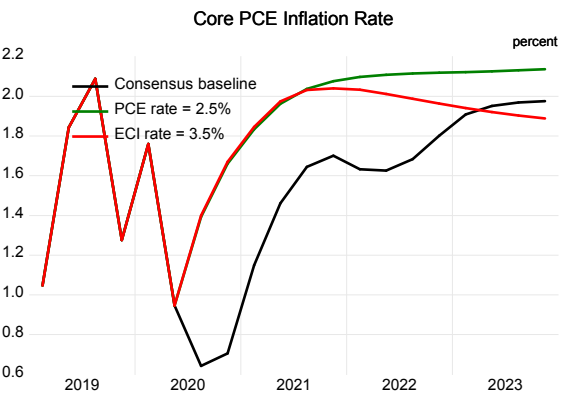

### 13. Macroeconomic Effects of Different Monetary Policy Reaction Functions
(VAR Expectations; Policy = rrftay)
(ZLB and Thresholds Imposed)

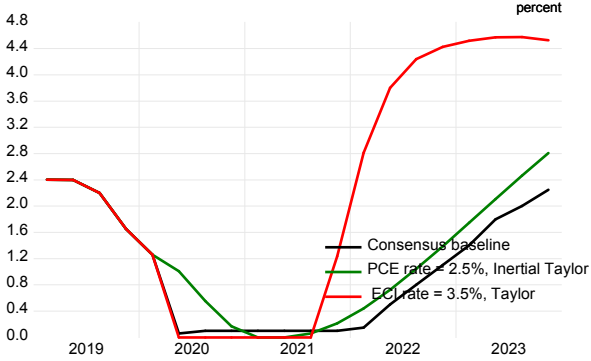
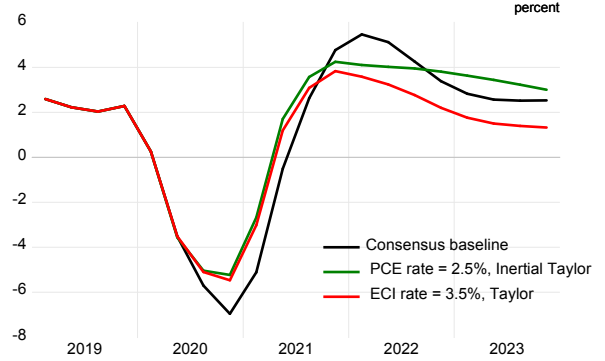
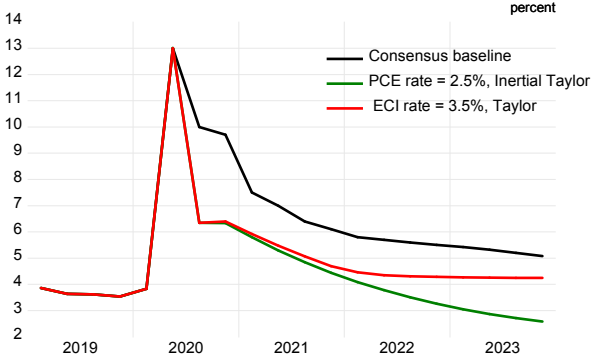
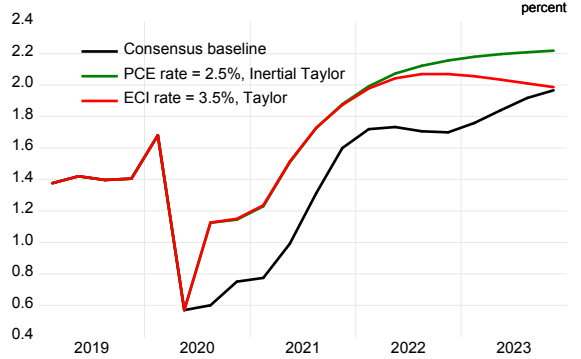
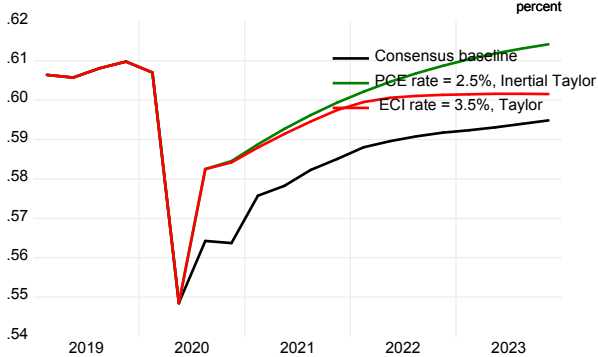
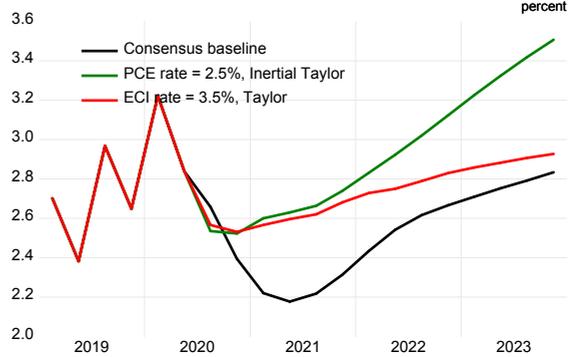
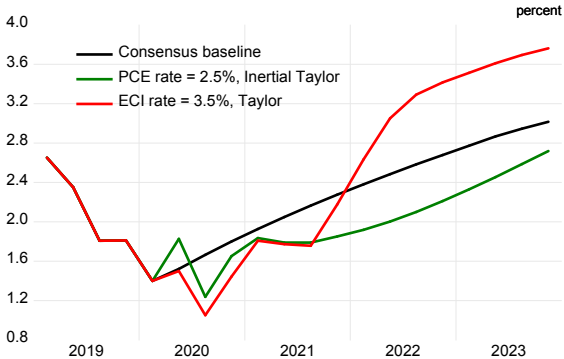
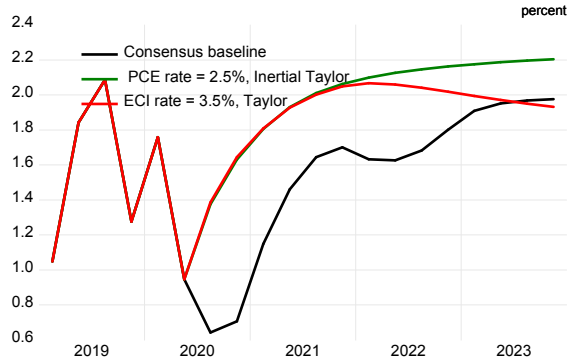

14. Macroeconomic Effects of a Modified Taylor Rule with a negative AD Shock
(VAR Expectations; Policy = rrftay)
(ZLB and Thresholds Imposed)

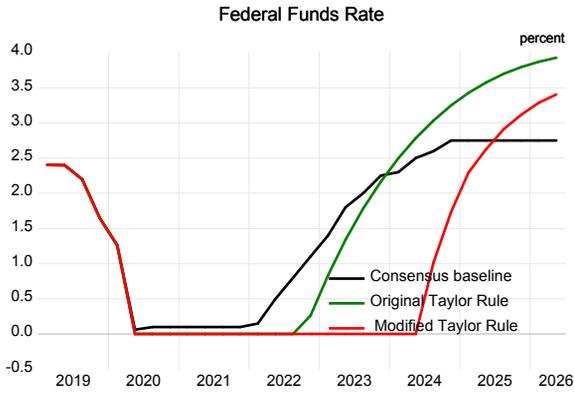
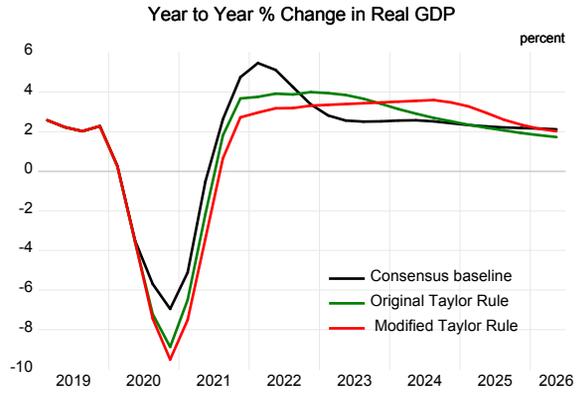
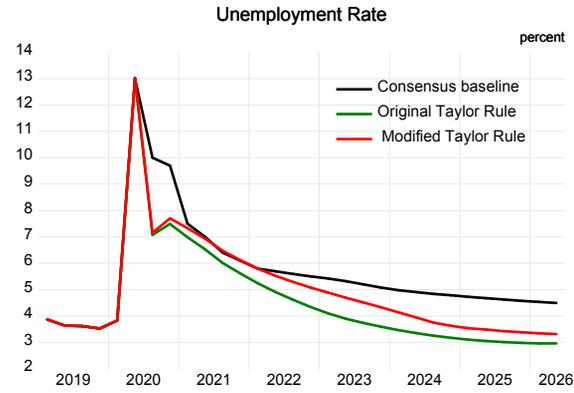
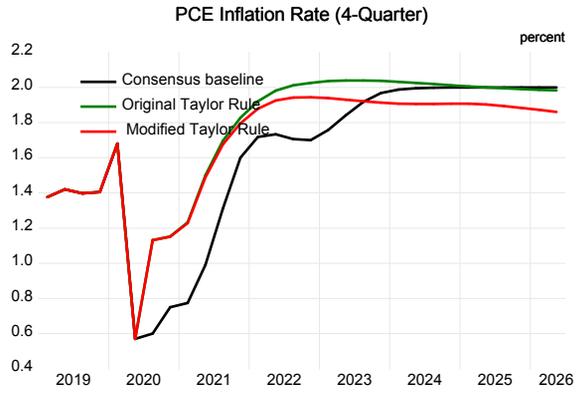
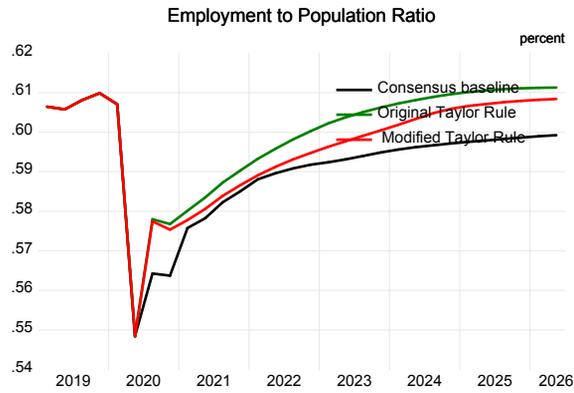
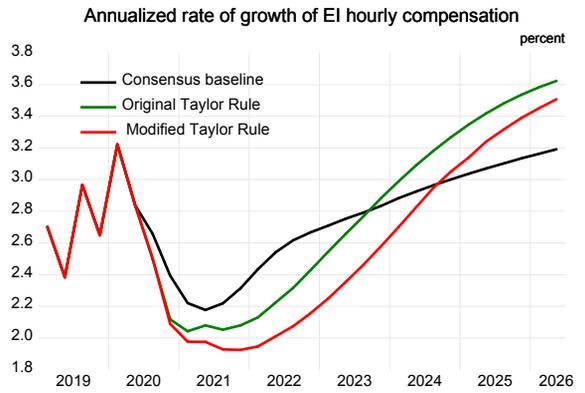
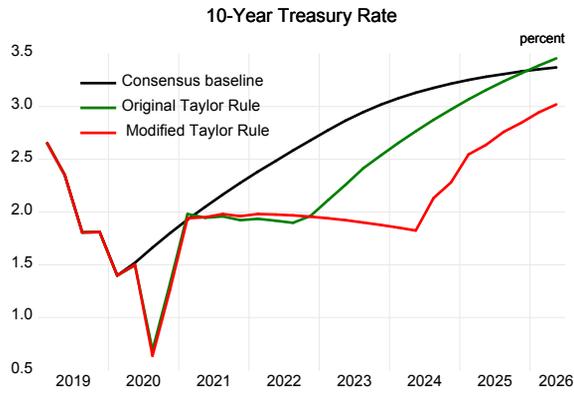
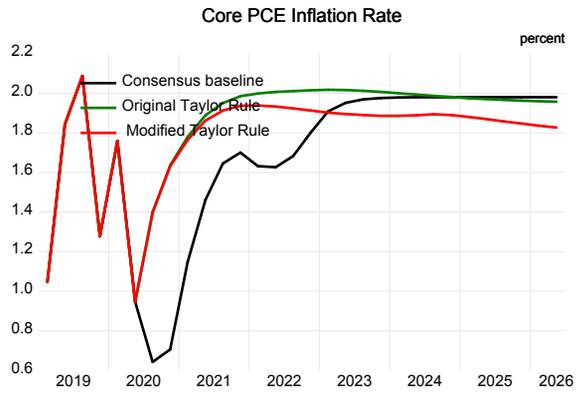

15. **Macroeconomic Effects of a negative AD Shock and Rise in the Federal Government expenditures**
(VAR Expectations; Policy = rrftay)
(ZLB and Thresholds Imposed)

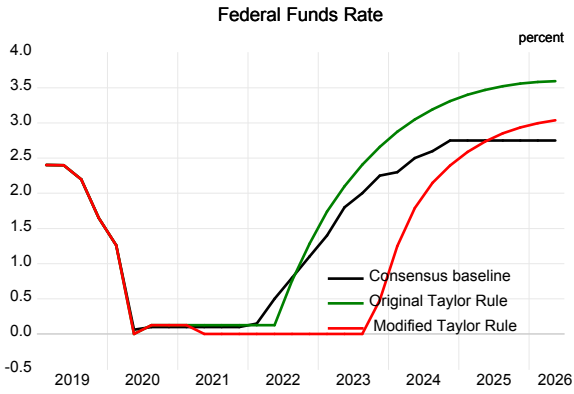
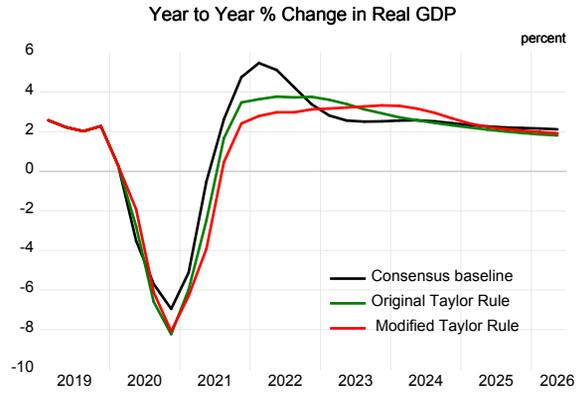
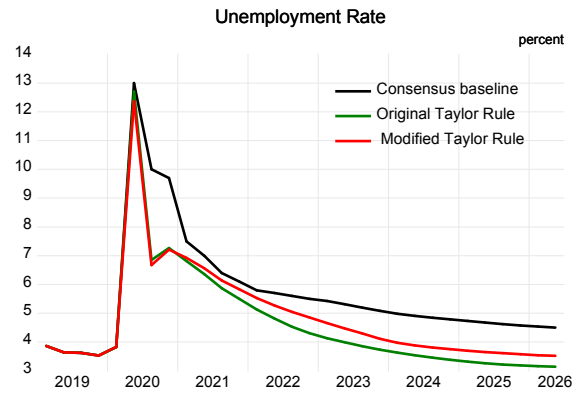
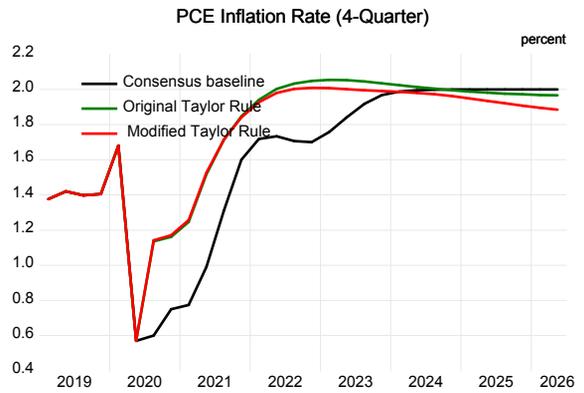
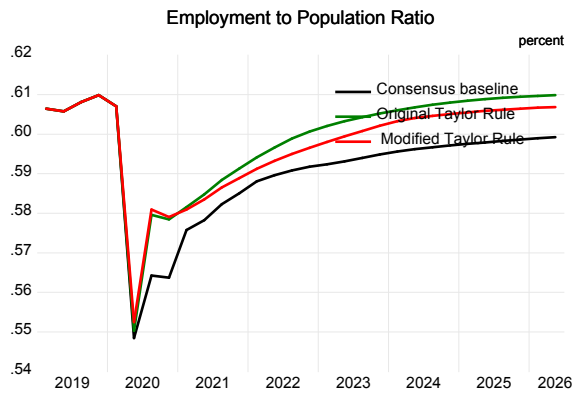
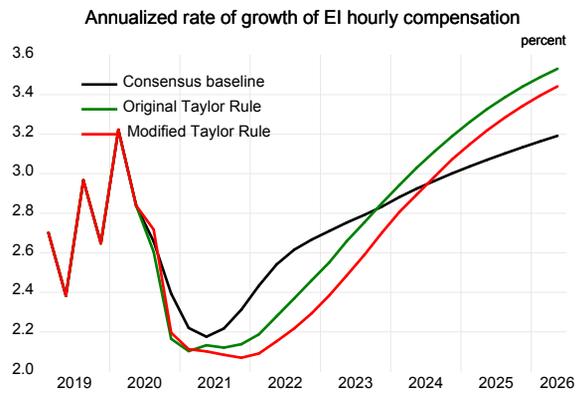
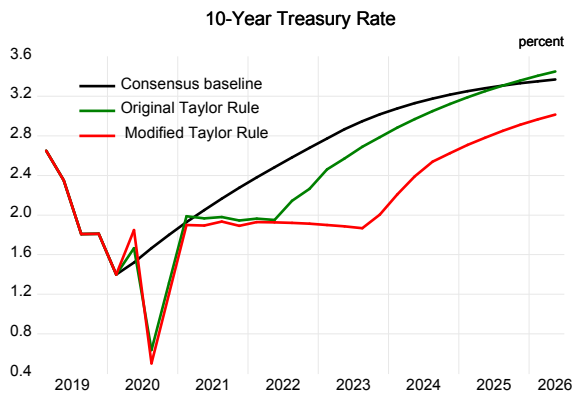
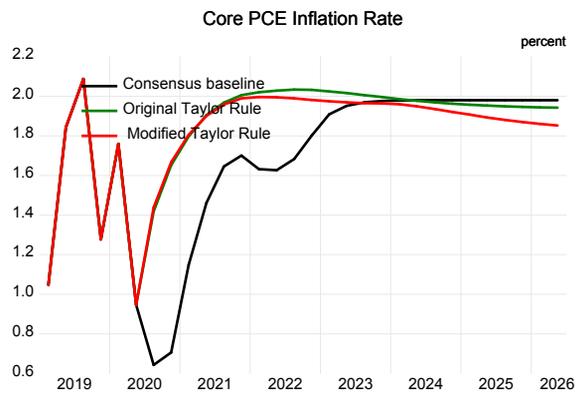